\documentclass[a4paper]{article}
\pdfoutput=1

\usepackage[utf8]{inputenc}
\usepackage[T1]{fontenc}
\usepackage[english]{babel}

\usepackage[pdftex,breaklinks=true,bookmarks=true,pdfborder={0 0 0},colorlinks=false]{hyperref}

\usepackage{listings}
\usepackage{amsfonts}
\usepackage{amsmath}
\usepackage{cleveref}
\usepackage{graphicx}
\usepackage[caption=false]{subfig}

\newcommand{\ie}{i.e.~}
\newcommand{\eg}{e.g.~}
\newcommand{\cf}{cf.~}
\newcommand{\vs}{vs.~}

\newcommand{\snet}{\textsc{S-Net}}
\newcommand{\spnet}{\textsc{S$^+$Net}}

\usepackage{enumitem}
\setlist{nolistsep}
\usepackage{array}
\newcommand*{\vcenteredhbox}[1]{\begingroup
\setbox0=\hbox{#1}\parbox{\wd0}{\box0}\endgroup}

\begin{document}

\author{R. Poss, M. Verstraaten, F. Penczek, C. Grelck, R. Kirner,
  A. Shafarenko\\University of Amsterdam, The Netherlands\\University
  of Hertfordshire, United Kingdom}
\title{\spnet{}: extending functional coordination \\ with
  extra-functional semantics}

\maketitle

\begin{abstract}
  This technical report introduces \spnet{}, a compositional
  coordination language for streaming networks with extra-functional
  semantics. Compositionality simplifies the specification of complex
  parallel and distributed applications; extra-functional semantics allow the application designer
  to reason about \emph{and} control resource usage, performance and
  fault handling. The key feature of \spnet{} is that functional and
  extra-functional semantics are defined orthogonally from each
  other. \spnet{} can be seen as a simultaneous simplification and
  extension of the existing coordination language \snet{}, that gives control of extra-functional
 behavior to the \snet{} programmer. \spnet{} can also be seen as
  a transitional research step between \snet{} and AstraKahn, another
  coordination language currently being designed at the University of
  Hertfordshire. In contrast with AstraKahn which constitutes a
  re-design from the ground up, \spnet{} preserves the
  basic operational semantics of \snet{} and thus provides an
  incremental introduction of extra-functional control in an existing
  language.
\end{abstract}

\clearpage

\setcounter{tocdepth}{2}
\tableofcontents

\clearpage

\section{Introduction}

\spnet{} provides a high level, declarative coordination language
based on concepts borrowed from stream processing. \spnet{} is based
on \snet{}: a coordination language whose notation describes
explicitly the data dependencies in a computation. As data movement is
then exposed in the language semantics, mapping and managing the
application to parallel platforms becomes simpler. \snet{} is
specified in~\cite{shafarenko.09.apc,penczek.10.snet} and has been
reported on in many published works,
including~\cite{grelck.08.lncs,grelck.08.ppl,grelck.10.ijpp,grelck.10.damp,holzenspies.10,penczek.10,grelck.11.ipdpsw,penczek.10.pcs,verstraaten.12,verstraaten.11}.

A key feature of \snet{} is its focus on \emph{declarative
  specifications}. The notation declares an intent of functional
composition of primitive networks into more complex applications.  The
management of execution, including automatic parallelisation and
automatic concurrent interleaving of activities, is fully delegated to a
run-time system in software.

\spnet{} answers two general concerns that have been revealed through the past use of
\snet{} with industrial applications.

For one, it was believed originally that the \snet{} technology ought
to be equipped with intelligence to automatically optimize application
execution, both during initial compilation and at run-time using
feedback loops. One of the goals of the EU-funded ADVANCE\footnote{\url{http://www.project-advance.eu}} project was
to attempt and demonstrate this. Unfortunately, one of the unescapable
lessons re-learned during ADVANCE is that at any level of intelligence
built into a system, there are some desirable optimizations that are
\emph{necessarily} out of reach from that system, although they could
be reachable by letting a human author refine the specification
manually~\cite{voeten.01.tdaes}.

Second, it was believed that the effectiveness of automated
optimizations would most often be superior to the human skill,
especially for large parallel systems, due to the complexity of the
systems involved. In practice, we observed a more nuanced
reality. \snet{} has revealed across its use cases that each
application has its own type of bottleneck, emerging as a consequence
of both over-constraining in the specification and run-time factors in
the implementation, such as hardware parameters or implementation
quirks of the software run-time
system~\cite{mckenzie.13.fdcoma}. Meanwhile, human operators equipped
with high-level monitoring and profiling tools have proved well-able
to understand these bottlenecks and subsequently restructure
applications to avoid them, often \emph{at a fraction of the cost that
  would be otherwise necessary to implement a new optimization} able
to recognize and handle the bottlenecks automatically.

This situation has motivated a new perspective on coordination, which
we attempt to capture with \spnet{}. On the one hand, general
principles of software engineering make it desirable to keep the
functional part of a specification devoid of operational
semantics. This
enables the automatically derivation of proofs of correctness
(\eg through the type system) and simplifies the mental model used
programmers during the initial phase of specification. On the other
hand, the coordination layer must provide tools to both
\emph{describe} the actual behavior of an implementation at run-time
(inspectability), and \emph{optionally prescribe} or constrain the
run-time behavior after a human operator has determined that the
additional prescriptions are desirable (adaptability). These
requirements call for a two-layered system, where functional and
extra-functional aspects of an application co-exist side-by-side and
can be used orthogonally from each other.

With this background requirement in place, the question remains of
what tools to place in the extra-functional toolbox. Following the
design guidelines of \snet{}, \spnet{} strives to provide
\emph{composable} specification operators that each determine an
orthogonal aspect of the system.

This following technical report thus describes \spnet{} and its
extra-functional contributions to \snet{}. In the process
of defining \spnet{}, the authors also took the opportunity
to recognize shortcomings in the functional part of \snet{}; the
functional core of \spnet{} is comparatively more simple and
its primitives are more orthogonal than \snet{}'s.

The report is organized as follows. In \cref{sec:fun} we present the
part of \spnet{}'s dedicated to functional specifications.  One of the
new functional concepts introduced in \spnet{} is the ``transducer''
component, subsequently further described in \cref{sec:translang}. In
\cref{sec:xfun} then presents \spnet{}'s extra-functional constructs
and their semantics. We finally provide a more in-depth discussion of the
differences between \snet{} and \spnet{} in \cref{sec:relsnet}.

\section{Functional specifications}\label{sec:fun}

The functional core of \spnet{} uses only two \emph{primitive
  components} and defines component network composition using
\emph{network combinators} over them.  The elementary components are
\emph{boxes} and \emph{transducers}, described in
\cref{sec:prim}. Boxes abstract entire programs provided externally to
\spnet{}. Transducers enable expressing simple computation and
synchronization constructs within the \spnet{} language. Two binary
combinators assemble heterogeneous composite networks, and three unary
combinators define more complex functional behaviors for their network
operand. The combinators are detailed in \cref{sec:funcombis}.

The interconnect between components appears as if each component had a
single input stream and output stream, that is, the input/output
events of a single component are observed in some total order. Each
event on such a stream is called a \emph{record}. Records are
represented as sets of label-value pairs.  Labels are called
\emph{field names} and values are called \emph{fields}. Fields are
either integer \emph{scalars}, or \emph{references} to data structures
from the box language domain.  References are opaque to coordination,
although \spnet{} and box languages must cooperate to marshall data to
and from streams of bytes suitable for communication over network
channels.

\subsection{Notations}

We present the composition constructs of \spnet{} in the following
sections.

As in \snet{}, record types are noted using curly brackets around the set of field
names, for example \lstinline!{a,b}! which is equivalent to
\lstinline!{b,a}!. Furthermore, \spnet{} standardizes the following notations:
\begin{itemize}
\item \emph{source notation}, \eg ``\texttt{A..B}'' which provides
   a standard syntax to enter specifications into the \spnet{} system;
\item \emph{algebraic notation}, \eg $\mathrm{C}(A, B)$, which
   provides a symbolic representation of input specifications, or
   specifications after partial automatic transformation by a \spnet{} implementation;
\item \emph{graphical notation},
  to visualize specifications and provide a high-level intuition.
\end{itemize}
The examples hereafter use either algebraic, source or
graphical notation, depending on which best illustrates a concept at hand.

\subsection{Primitive networks}\label{sec:prim}

\begin{table}
\centering
\begin{tabular}{p{.15\textwidth}cc>{\ttfamily}c}
Name & Graphical representation & Algebraic & {\normalfont Source notation} \\
\hline
Box & \vcenteredhbox{\includegraphics[scale=.4]{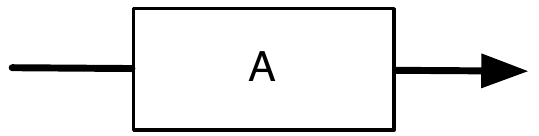}} & $\underset{\mathcal{B}\rightarrow\mathcal{N}}{\mathsf{Box}(A)}$ & box A ... \\

Transducer & \vcenteredhbox{\includegraphics[scale=.4]{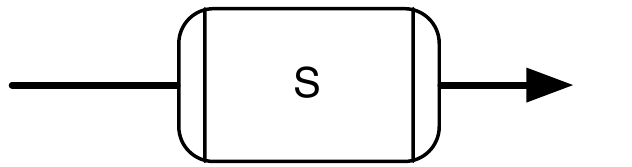}} & $\underset{\mathcal{S}\rightarrow\mathcal{N}}{\mathsf{Transduce}(S)}$ & \verb+[|+$S$\verb+|]+ \\
\end{tabular}
\caption{Notations for \spnet{}'s primitive networks.}\label{tab:prim}
\end{table}

\spnet{} provides two primitive networks, from which all
specifications are derived: \emph{boxes}, which capture components
developed externally, and \emph{transducers}, which are stateful boxes
whose behavior is implemented using a simple expression language
within \spnet{}.  The corresponding notations are given in
\cref{tab:prim}.

\subsubsection{Boxes: stateless transformations}\label{sec:boxes}

The main elementary functional construct is the encapsulation of an
entire program, called component or \emph{box}, able to work
asynchronously on a stream of input and producing a stream of output.

Both imperative and declarative languages qualify as box
implementation languages. The box language infrastructure around each
component must offer clear primitives to set up a box, provide it with
input, save or checkpoint its management state (\eg heap managers,
debugging metadata) and tear down the environment. A box
implementation should be resource-agnostic, that is, able to perform
its transformation regardless of the hardware resources seleted for it
by the \spnet{} coordination layer. In its simplest form, a box
encapsulates a side-effect free, stateless function implemented in a
standard programming language like C or C++.

A box can be described as a function from
its input stream to its output stream. Boxes must
appear functionally pure, that is, the output from one input
record can be fully computed using that input record only,
and a box terminates after it produces its last output.
A box is characterized by a \emph{box signature}: a mapping
from an input type to a disjunction of output types. For example,
\begin{center}
\texttt{box foo ((a, b) -> (c) | (c,d));}
\end{center}
declares a box \texttt{foo} that expects records with two fields
labelled \texttt{a} and \texttt{b}. The box, when activated, responds
with zero or more records that either have only one field labelled
\texttt{c}, or two fields labelled \texttt{c} and \texttt{d}.

The set of field names naturaly induces a \emph{type signature} for
every stream-to-stream transformation\footnote{
General type signatures use set notation for record types,
with curly brackets. For box
signatures, order matters: the box implementation might only support
positional arguments and record fields are then provided at run-time
in the order specified. This is why box signatures use parentheses
instead of curly brackets.}. Type compatibility and
subtyping are determined by set inclusion. For example, the box
\texttt{foo} above would accept a record with type
\lstinline!{b,a,c}! as input, but not \lstinline!{a}!  nor
\lstinline!{b,c,e}!.  Subtyping on input types of boxes raises the
question of what happens to the excess fields.  \spnet{} defines
\emph{flow inheritance}, whereby excess fields from incoming records
are not just ignored in the input record of a network entity, but are
also attached to any outgoing record produced by it in response to
that record.  Subtyping and flow inheritance prove to be indispensable
features when it comes to combine boxes designed in isolation into a
larger application.

\subsubsection{Transducers: stateful, finite synchronizers}\label{sec:trans}

A box, already described in the previous section, is stateless and can
only split a record into parts.  The transducer construct expresses
the complementary operation: merging two or more records into one,
possibly performing computations on them.  Transducers are defined as
finite state machines\footnote{
  theoretical construct of the same name, described
  in~\cite[Chap.~4]{sakarovitch.09}, of which they are a practical
  application.}; the number of different states is kept finite so that
typing and correctness can be decided statically, and so that liveness
and future state behavior can be predicted at run-time.

All specifications for transducers specify a finite set of states,
conditions for transitions between them, and actions to carry out upon
state changes. Actions include capturing (part of) the input using
\emph{hold variables} and/or constructing records to emit on the
output stream.  Hold variables can capture one record each, and have
full/empty semantics: they can only be assigned when empty, and read
or reset to the empty state when full.

Transitions can be conditional on input records, but conditions cannot involve
previously captured input. Consequently, the type signature of the
transducer can be determined statically for every state, as well as
whether the hold variables stay consistent, \ie each hold variable is
not set when full and not reset when empty.

Any input record that arrives to a transducer and not accepted by a
transition from the current state is output, unchanged, by the
transducer. A transducer is said to be \emph{inactive} whenever it
is currently in its initial state and all its hold variables are
empty; it \emph{terminates} when it reaches a state with no outgoing
transition.

More details on the transducer sub-language are given in \cref{sec:translang}.

\subsubsection{Multiple input, multiple output and tags}

A distinguishing feature of \spnet{} is that it neither introduces
streams as explicit objects nor does it defines network connectivity
through explicit wiring.  Instead, it uses algebraic formulae, described
below in \cref{sec:funcombis}, to
compose streaming networks.  The restriction of type signatures to a
single logical input and a single logical output stream (SISO) is essential
for this. However, this is not to say that \spnet{} actually
implements logical streams using a single communication channel; MIMO
specifications are also possible. For this, records may be
additionally be marked using a single \emph{tag}, noted between angle
brackets, for example $\langle t\rangle$. The inclusion of a tag in a
component signature changes the type equivalence to only match records
with exactly the same tag.  Therefore tags in one component's output
can only be matched to components with the same tags in their declared
input type, and an implementation can route this communication through
dedicated channels. Tags may also be used as scalars to distinguish
between sub-streams.

%

\subsection{Basic operational model for state management}
\label{sec:liveness}

Stateful constructs like the transducer evolve through a
lifecycle at run-time. Other functional combinators define below
also define implicit state. State necessarily occupies
space at run-time and breaks referential transparency, and
thus defines \emph{objects} at run-time that must be managed.

The basic properties of these objects are defined below; the
rest of the \spnet{} specification refers to these properties
and determines how they are managed by a run-time system.

To start, any stateful construct
goes through the following phases:
\begin{enumerate}
\item when initially instantiated, it is \emph{inactive};
\item during its lifetime, it may go through one or more
  \emph{activity cycles}, during which it is stateful, interleaved
  by inactivity periods;
\item it may reach a state past which it either always behave like the
  identity function, or never processes input records again, at which
  point it is said to be \emph{terminated}.
\end{enumerate}

A (sub-)network is said to be \emph{live} as long as it is not
terminated.  Note that a network can be both active and terminated.
For example, a transducer is known to be terminated whenever it
reaches a state with no possible transition. However, at that point it
may still have non-empty hold variables. In accordance to previous
work in process management, we call \emph{dead} those networks which
are both terminated and inactive; and \emph{zombie} those which are
terminated but still active (retain state).

When proven at run-time, the death property enables the simplification
of the process network by eliminating replicas or run-time state that
have become unnecessary, \ie \emph{garbage
collection}~\cite{grelck.11.ipdpsw}. Depending on the application
domain, the implementation may also permit automatic garbage
collection of zombie networks.

In this context, the lifecycle of primitive networks deserves
attention.  It is possible to consider a box as a network that starts
inactive, then becomes repeatedly active for each successive input
record. Because it is stateless, it is also possible to consider it as
a network which terminates after each successive input record, then
instantiated anew for the next input record. Similarly, whenever a
live transducer becomes inactive, either it can be re-activated for
the next input record or terminated and reinstantiated anew. This
duality is functionally neutral; however, which approach is taken in
an implementation has extra-functional consequences on latency and
throughput, and is thus observable and controllable in \spnet{}, via
the use of projections (\cf \cref{sec:projs,sec:projmotiv}).

\subsection{Functional combinators and composite networks}\label{sec:funcombis}

\spnet{} provides the following functional combinators, which
compose networks:
\begin{itemize}
\item finite \emph{composition} of two or more networks, to define pipelines;
\item finite \emph{selection} between two or more networks, to
 define routing between alternative processing pipelines;
\item arbitrarily deep \emph{ordered replication composition}, which
  dynamically replicates a network and composes the replicas in
  a deterministically ordered pipeline;
\item arbitrarily deep \emph{unordered replication composition}, which
  dynamically replicates a network and composes the replicas in
  an non-deterministically ordered pipeline.
\end{itemize}

Both simple selection and the replication
operators introduce processing concurrency, that is, non-determinism
in the order records are visible in the logical output stream. To
force reordering when needed, \spnet{} also provides a
\emph{reordering} combinator which ensures that output
records are visible in the same order as the input record
that caused them.

\begin{table}
\centering
\begin{tabular}{p{.15\textwidth}cc>{\ttfamily}p{.2\textwidth}}
Name & Graphical representation & Algebraic & {\normalfont Source notation} \\
\hline
Composition & \vcenteredhbox{\includegraphics[scale=.4]{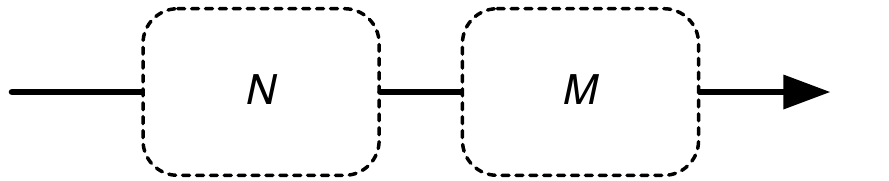}} & $\underset{\mathcal{N^*}\rightarrow\mathcal{N}}{\mathrm{C}(...)}, \text{ eg. } \mathrm{C}(N,M)$ & $N$..$M$ \newline \normalfont ``dot dot'' \\

Selection & \vcenteredhbox{\includegraphics[scale=.4]{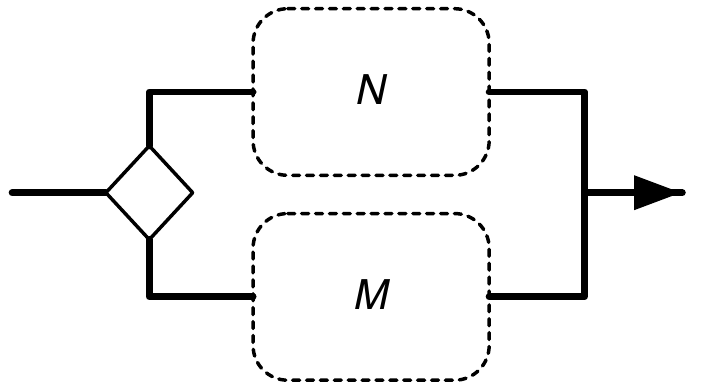}} & $\underset{\mathcal{N^*}\rightarrow\mathcal{N}}{\mathrm{S}(...)}, \text{ eg. } \mathrm{S}(N,M)$ & $N$|$M$ \newline \normalfont ``or'' \\

Ordered replication composition & \vcenteredhbox{\includegraphics[scale=.4]{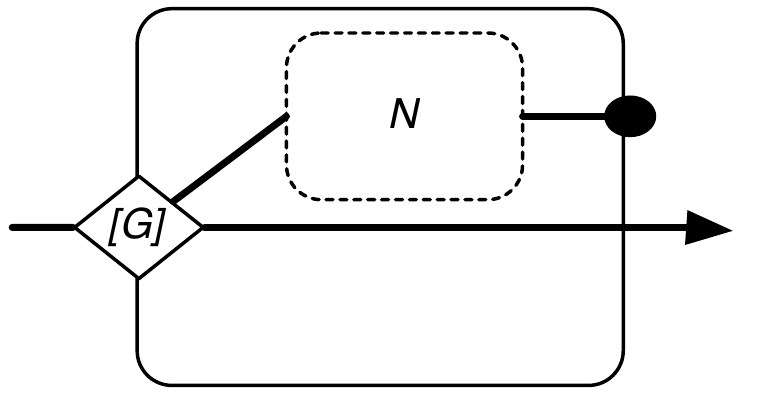}} & $\underset{\mathcal{G}\times\mathcal{N}\rightarrow\mathcal{N}}{\mathrm{C^*_{G}}(N)}$ & $N$*$G$ \newline \normalfont ``star'' \\

Unordered replication composition & \vcenteredhbox{\includegraphics[scale=.4]{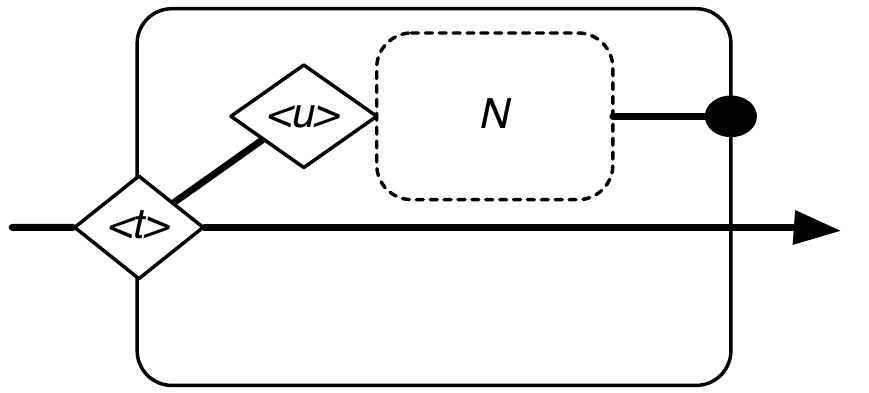}} & $\underset{\mathcal{T}^2\times\mathcal{N}\rightarrow\mathcal{N}}{\mathrm{C^{!}_{\langle t\rangle,\langle u\rangle}}(N)}$ & $N$!*<$t$><$u$> \newline \normalfont ``blink star'' \\

Reordering & \vcenteredhbox{\includegraphics[scale=.4]{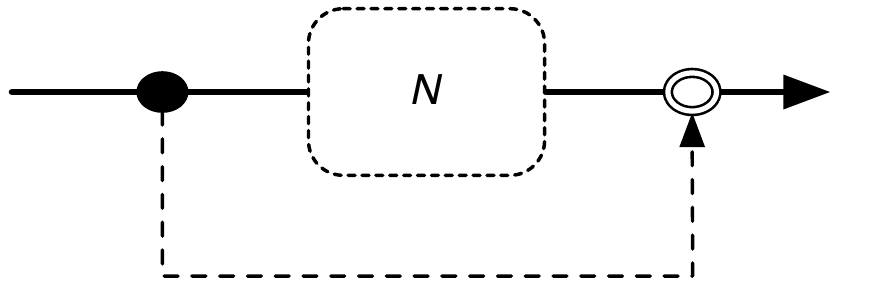}} & $\underset{\mathcal{N}\rightarrow\mathcal{N}}{\mathrm{R}(N)}$ & ?$N$\# \\
\end{tabular}
\caption{Notations for \spnet{}'s functional combinators.}\label{tab:fequiv}
\end{table}

The corresponding notations are given  in \cref{tab:fequiv}. The following sections
introduces each combinator in more details.

\subsubsection{Simple composition and selection}

Composition ($\mathrm{C}(A,B)$, or ``\texttt{A..B}'' in source)
constructs a new network where the logical streams are connected in
series. The composite network can be thought as performing the
function $B\circ A$ concurrently over each input record to produce output
records. This form of composition preserves ordering of the input
stream in the output stream.

Selection ($\mathrm{S}(A,B)$, or ``\texttt{A|B}'' in source)
constructs a new network where each input record is routed to one of
the branches. Which route is taken is determined by best match on the
input type of the alternative networks. The composite network can be
thus thought as performing a type-based functional choice between $A$
or $B$. Moreover, selection introduces stream concurrency between the
alternatives: when two successive records $r_a$ and $r_b$ are
presented on the input stream, to be routed to $A$ and $B$, $B$'s
output for $r_b$ may appear interleaved in any way with $A$'s output
for $r_a$ on the output stream.
Ambiguous selections are resolved in the specification order\footnote{
This selection determinism is a divergence from the approach
taken with  \snet{}'s ``parallel composition'' construct; we
discuss this in \cref{sec:relsnetfun,sec:ersel}.}. For
example with $\mathrm{S}(A,B)$, if $A$ accepts $\{a\}$ and $B$ accepts
$\{b\}$, a record with type $\{a,b\}$ will be routed to
$A$.

At any point during execution, a selection is active if at least one
of its branches is active; when it is terminated, it terminates also
all the branches.  We define by extension the \emph{dynamic liveness
  arity} and \emph{dynamic activity arity} of a composite, which is
the current number of live and active instances of the inner networks
at run-time, respectively. Dynamic liveness arity is intuitively
associated with passive spatial complexity, implied by the cost of
maintaining the network instances in the run-time environment; dynamic activity
arity is intuitively associated with active spatial complexity,
implied by the cost of actual accesses to the instances' state.

The logical output stream of the composite network is the fusion
of the concurrent sub-streams after processing by the replicas. As
with simple selection, fusion is non-deterministic and sub-streams
may appear interleaved.





\subsubsection{Reordering}

If the non-determinism in the output order of selection is not
desired, it is possible to encapsulate a network $N$ in the reordering
combinator which reintroduces the input order, noted
$\mathrm{R}(N)$. For example, with $\mathrm{R}\circ\mathrm{S}(A,B)$,
or ``\texttt{?A|B\#}'' in source form, when presented with successive
records $r_a$ and $r_b$, $A$'s last output for $r_a$ will be observed
on the output stream of the composite network before $B$'s first
output for $r_b$.

At any point during execution $\mathrm{R}(N)$ is active
if either $N$ is currently active or if the composite
network is currently holding records for reordering. It terminates
$N$ when terminated itself.

\subsubsection{Ordered replication composition}

The ordered replication composition over $N$, noted \texttt{N!G} in
source form or $\mathrm{C}^*_{G}(N)$ in algebraic form, defines
a composite network whose functional semantics are defined as follows:
\begin{itemize}
\item $G$ is a \emph{guard pattern}, which expresses record types
  together with an optional field predicate;
\item it processes records through an ordered composition of replicas of $N$ as long
  as records do \emph{not} match the guard pattern.
\end{itemize}

In other words, ordered replication composition can be thought
of defining an infinitely enumerable set of replicas $\{\lfloor N\rfloor_i\mid i \in \mathbb{N}\}$ and
processing each input record via the functional definition
\[
\begin{aligned}
\lfloor \mathrm{C}^*_{G}(N)\rfloor &= \mathrm{C}^*_{G}(\lfloor N\rfloor_0) \\
\mathrm{C}^*_{G}(\lfloor N\rfloor_i) (r) &= \begin{cases}
r & \text{if } r \text{ matches } G \\
\mathrm{C}(\lfloor N\rfloor_i, \mathrm{C}^*_{G}(\lfloor N\rfloor_{i+1}) (r)  & \text{otherwise} \\
\end{cases}
\end{aligned}
\]

\begin{figure}
\centering
\includegraphics[scale=.4]{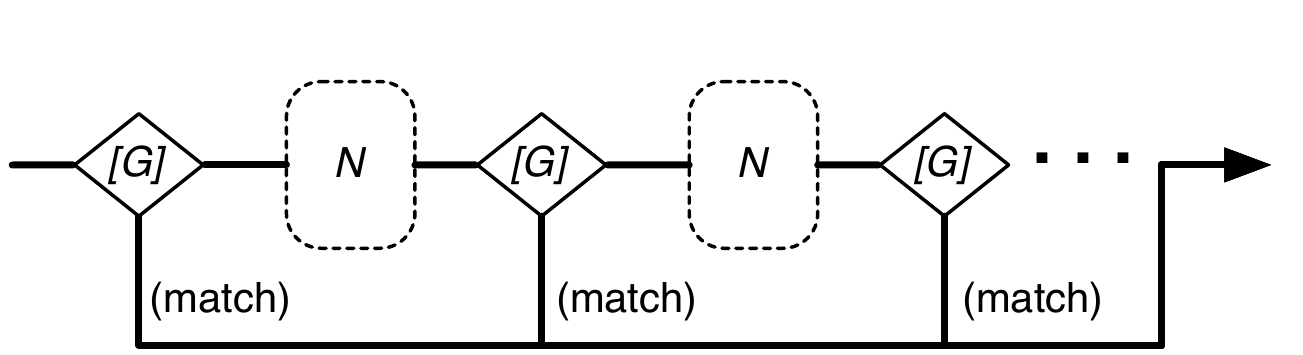}
\caption{Dynamic unfolding of $\mathrm{C}^*_G(N)$.}\label{fig:starunfold}
\end{figure}

The corresponding dynamic unfolding is illustrated in \cref{fig:starunfold};
it implements guarded functional
recursion. This is the alternative proposed by \spnet{}
to cycles in specification graphs to define repeated behavior.

Like with selection, activity is defined inductively from
the corresponding state of the inner replicas, and termination of the
composition implies termination of the replicas. Dynamic liveness and
activity arities are also defined transparently. In addition,
replication composition defines \emph{dynamic depth} to be the number
replicas involved in the linear chain that are not yet
terminated. Dynamic depth is intuitively associated with the time
complexity of processing, since any terminated replica is
neutral w.r.t composition.

An implementation is expected to remove terminated replicas dynamically
from the composition chain, \ie keep the dynamic liveness arity synchronized
with the dynamic depth.


\subsubsection{Unordered replication composition}\label{sec:urc}

The unordered replication composition is intended to capture the
non-deterministically ordered sequence of transformations on a shared
data structures between computation agents. For example, it can be
used to express the transformation of a graph data structure as an
non-deterministically ordered set of concurrent
subgraph-to-subgraph transformations.  The intent is to let the
\spnet{} implementation schedule the concurrent activities either in
parallel using either locking or speculation and transactional storage to
guarantee the linear order (\cf \cref{sec:aumotiv}).

The functional semantics are defined as follows.  The construct
\texttt{N!*<r><p>} in source form or $\mathrm{C}^!_{\langle
  r\rangle,\langle p\rangle}(N)$ in algebraic form, defines a
composite network. The first tag $\langle r\rangle$ marks
\emph{constructor records} and the second tag  $\langle p\rangle$
marks \emph{payload records}.  Any contiguous sub-sequences of
constructor records is called a \emph{constructor sub-sequence}. A
single constructor sub-sequence (of zero or more constructors)
followed by one payload record is called a \emph{processing
  sub-sequence}.  Any record that are neither constructors or
payloads, called \emph{pass-through records}, are forwarded to the
output stream unchanged.

In each processing sub-sequence of the input stream, the behavior is defined as follows:
\begin{itemize}
\item the constructors are collected in a list;
\item if there are no constructors, the payload is passed through; otherwise
\item one constructor is picked \emph{at a non-deterministic position} of the list and removed from the list;
\item the picked constructor and the payload are presented as input
   to  a new (fresh) replica of the inner network;
\item any constructor produced by the inner network is appended to the constructor list;
\item the resulting constructor list and the payload produced by the inner network are
  processed by a new instance of
  $\mathrm{C}^!_{\langle r\rangle,\langle p\rangle}(N)$;
\item the original replica is terminated.
\end{itemize}

\begin{figure}
\centering
\includegraphics[scale=.4]{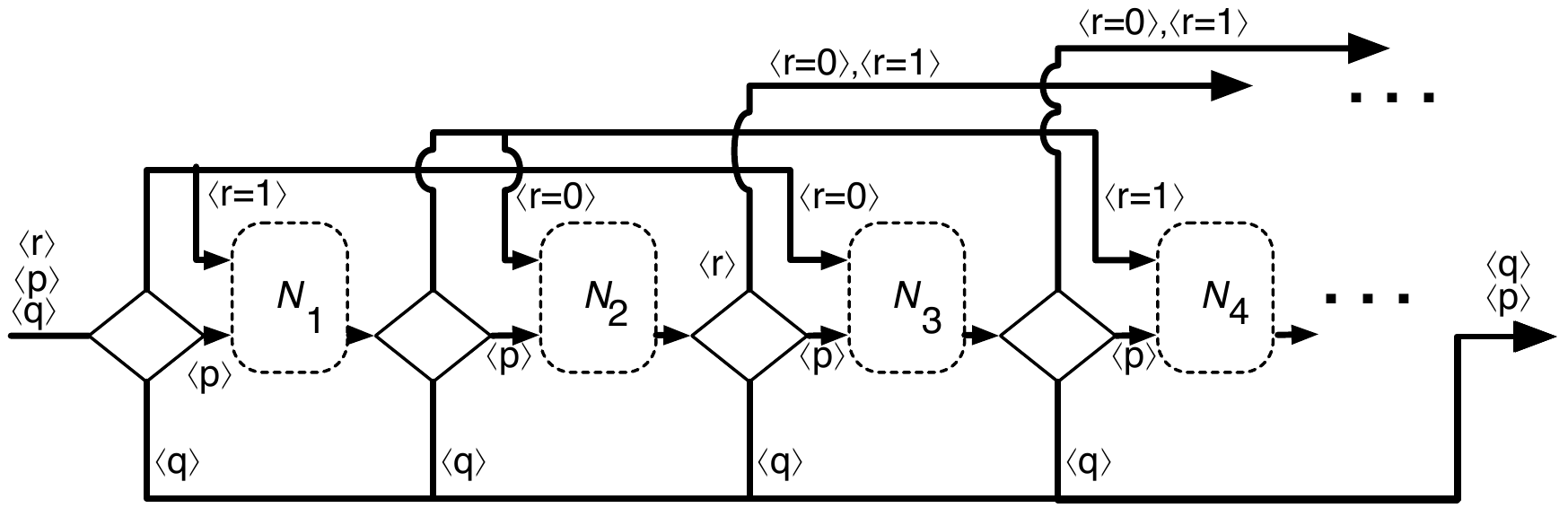}
\caption{Dynamic unfolding of $\mathrm{C}^!_{\langle r\rangle,\langle p\rangle}(N)$.}\label{fig:bstarunfold}
\end{figure}

An example dynamic unfolding is illustrated in
\cref{fig:bstarunfold}. In this example, each replica of the inner
network accepts one $\{\langle r\rangle\}$ and one $\{\langle
p\rangle\}$ record. Each replica also emits conditionally two new
records $\{\langle r=0\rangle\}$ and $\{\langle r=1\rangle\}$ followed
by a new $\{\langle p\rangle\}$ record, possibly interleaved with
$\langle q\rangle$ records.  When the processing sub-sequence
$\{\langle r=0\rangle\}, \{\langle r=1\rangle\}, \{\langle p\rangle\}$
is presented on the input, the following happens. The first two
$\langle r\rangle$ records are accumulated in a list.  The 2nd
constructor is picked first ($\langle r=1\rangle$), a first replica
$N_1$ is created, and both the constructor and $\langle p\rangle$
record are presented to $N_1$. The $\langle r\rangle$ records produced
by $N_1$ are then appended to the list, which becomes $[\{\langle
  r=0\rangle\},\{\langle r=0\rangle\},\{\langle r=1\rangle\}]$.  Again
the 2nd constructor is picked first ($\langle r=0\rangle$ from $N_1$),
and defines a new replica $N_2$.  The $\langle p\rangle$ record output
by $N_1$ is then sent to $N_2$. The $\langle r\rangle$ records
produced by $N_2$ are appended to the list, then the 1st element is picked
from the remaining list ($\langle r=0\rangle$ from $N_2$). And so on.  Meanwhile,
all the interleaved $\langle q\rangle$ records are also forwarded to
the output stream.

All processing sub-sequences of the logical input stream are processed
concurrently, by distinct processing sets of replicas.  In each
processing set, any given replica is only used once, over exactly one
constructor and one payload record. The replica is terminated after
it has processed the input payload and optionally output one new
payload. A single top-level processing sequence is said to
\emph{complete processing} when all replicas in its processing set
have terminated and no constructor records are left unprocessed.

As can be seen, this combinator combines two forms of non-determinism:
both concurrency between input processing sequences and concurrency
between constructor records at any level of the recursion, which
non-deterministically change the composition order of the remaining
replicas.

As with the previous forms of replication, activity
is defined inductively from the
corresponding state of replicas. When terminated, the replication
terminates its replicas. Dynamic liveness and activity arities
are defined as the total number of live/active replicas across all
processing sets. The dynamic depth of the network is is the maximum of
the dynamic depths of the individual processing sets. As with ordered
replication composition, an implementation is expected to remove terminated
replicas automatically from the run-time environment.

\section{Transducer language}
\label{sec:translang}

A box, already described in the previous section, is stateless and can
only split a record into parts. Moreover, a box is defined externally
to \spnet{} and its specification is not known at the level of
\spnet{}. In contrast, transducers provide a way to define simple
computations on records within the \spnet{} language itself. Like
boxes, once defined, transducers behave as primitive networks (\cf
\cref{sec:prim})

\subsection{Overview}

 The transducer construct expresses
the complementary operation: merging two or more records into one,
possibly performing computations on them.  Transducers are defined as
finite state machines\footnote{Transducers are named after the
  theoretical construct of the same name, described
  in~\cite[Chap.~4]{sakarovitch.09}, of which they are a practical
  application.}; the number of different states is kept finite so that
typing and correctness can be decided statically, and so that liveness
and future state behavior can be predicted at run-time.

All specifications for transducers specify a finite set of states,
conditions for transitions between them, and actions to carry out upon
state changes. Actions include capturing (part of) the input using
\emph{hold variables} and/or constructing records to emit on the
output stream.  Hold variables can capture one record each, and have
full/empty semantics: they can only be assigned when empty, and read
or reset to the empty state when full.

Transitions can be conditional on input records, but conditions cannot involve
previously captured input. Consequently, the type signature of the
transducer can be determined statically for every state, as well as
whether the hold variables stay consistent, \ie each hold variable is
not set when full and not reset when empty.

Any input record that arrives to a transducer and not accepted by a
transition from the current state is output, unchanged, by the
transducer. A transducer is said to be \emph{inactive} whenever it
is currently in its initial state and all its hold variables are
empty; it \emph{terminates} when it reaches a state with no outgoing transition.

\subsection{Specification language}

\spnet{} provides a comprehensive syntax for transducers, to keep
the specification short in simple cases and to factor regular behavior.
To start with, it is common to express housekeeping coordination tasks
that are stateless, for example duplicating records or performing
simple arithmetic on scalar fields. Transducers can be defined with
only one state and no captured input as follows:
\begin{center}
\texttt{ [| \{a, b, c\} -> [emit \{a=input.a,
      z=input.a, t=0\}; emit \{b, a=input.b, c=input.c+1\}] |] }
\end{center}
This transducer consumes records of type \{a,b,c\} and for each input creates two new records.  The first
output record has field \texttt{a} with the original value, field \texttt{z}
with the same value and a scalar \texttt{t} set to zero.  The second
record has fields \texttt{b} with the original value, \texttt{a} with
the same value as \texttt{b} and the scalar \texttt{c} incremented by
one.  A stateless transducer is also called ``filter'', and equivalent to the \snet{} construct of the same name.

To capture input, a specification must define hold variables.
For example, the transducer defined with
\begin{center}
\begin{verbatim}
[| var x;
 start: {a} -> [x := input] ha; {b} -> [x := input] hb;
 ha: {b} -> [emit input + x; reset x] start;
 hb: {a} -> [emit input + x; reset x] start;
|]
\end{verbatim}
\end{center}
has one hold variable \texttt{x}.  When in the \texttt{start} state, it
accepts either records of type $\{a\}$ or $\{b\}$, captures them in
\texttt{x} and changes to a state where it can accept a record of the
other type, combine it (using set union) and output the result, before
resetting to the initial state.

The record expression after the label must be an exact match
on the input record. The labels must match, and a scalar value, if specified,
must match the input's scalar as well; this uses first match if there are multiple guards in the same state. For example,
\texttt{[| \{a=9\} -> [emit \{a=0\}];
   \{a\}   -> [emit \{a=input.a+1\}]; |]}
 specifies a filter that increments  \texttt{a} modulo 10.

To enable inheritance, a transducer can accept records with more
fields than are matched in the guard, as follows:
\texttt{[| \{a\}+x -> [emit \{a=input.a+1\}+x] |]}.
This specifies a filter that accepts any input with at least
field \texttt{a} and produces as output a record with \texttt{a}
incremented, together with the remaining fields of the input.
By extension, a guard can omit the match pattern and only use a ``remainder'' part, for example:
\texttt{
[| a: x -> [emit x+\{even=1\}] b;
   b: x -> [emit x+\{odd=1\}]  a; |]}.
This transducer adds scalars \texttt{odd} or \texttt{even} to alternate input records, regardless
of their type.

It is also possible to manipulate state labels as integer values within
a fixed range, and use \emph{finite-size arrays} of hold variables
indexed by the state labels. For example:
\begin{center}
\begin{verbatim}
[| label [0..3]; var h[3];
   n=0..2/  n: x -> [h[n] := x] n+1;
            3: x -> [emit union(h)+x; reset h] 0; |]
\end{verbatim}
\end{center}
This specifies 4 states labeled from 0 to 3, and 3 hold variables. The transition prefix
``\texttt{n=0..2/}'' defines a \emph{label iterator} \texttt{n} and
causes the duplication of the remainder of the transition
specification for all specified values of \texttt{n}.  Therefore, this
transducer merges every successive 4 records into one, regardless of
type.

Label iterators can be used with finite-size arrays
of guards, too.  For example, the transducer
\begin{center}
\begin{verbatim}
[| label [0..3]; var h[3]; guard t[3] = {a},{b},{c};
   n=0..3/  n: t[n] -> [h[n] := input] n+1;
            4: {d} -> [emit union(h)+input; reset h] 0; |]
\end{verbatim}
\end{center}
 merges subsequences of $\{a\}$, $\{b\}$, $\{c\}$ and
 $\{d\}$ records, exactly in that order.

\begin{figure}
\centering
\subfloat[Specification]{
\begin{minipage}{.7\textwidth}
\scriptsize\ttfamily
[| guard t[3] = \{a\},\{b\},\{c\}; label s[3]; var h[3];

i=0..2 / \~{}s[i]: t[i] -> [h[i] := input] s[i];

   s[0..2]: -> [emit union(h); reset h] \~{}s[0..2]; |]
\end{minipage}
}

\subfloat[Specified transitions]{\includegraphics[scale=.4]{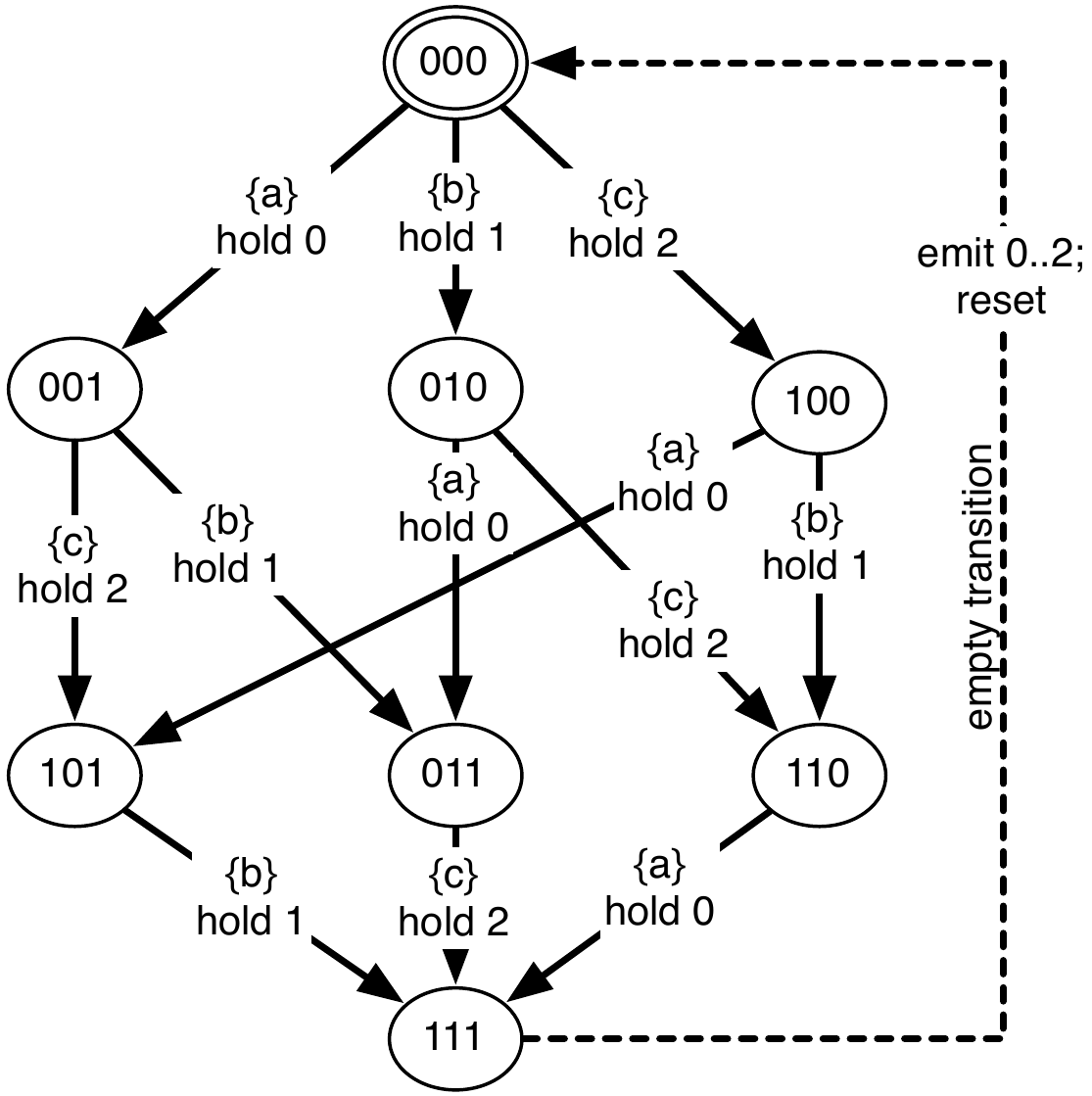}}\quad
\subfloat[Actual transitions]{\includegraphics[scale=.4]{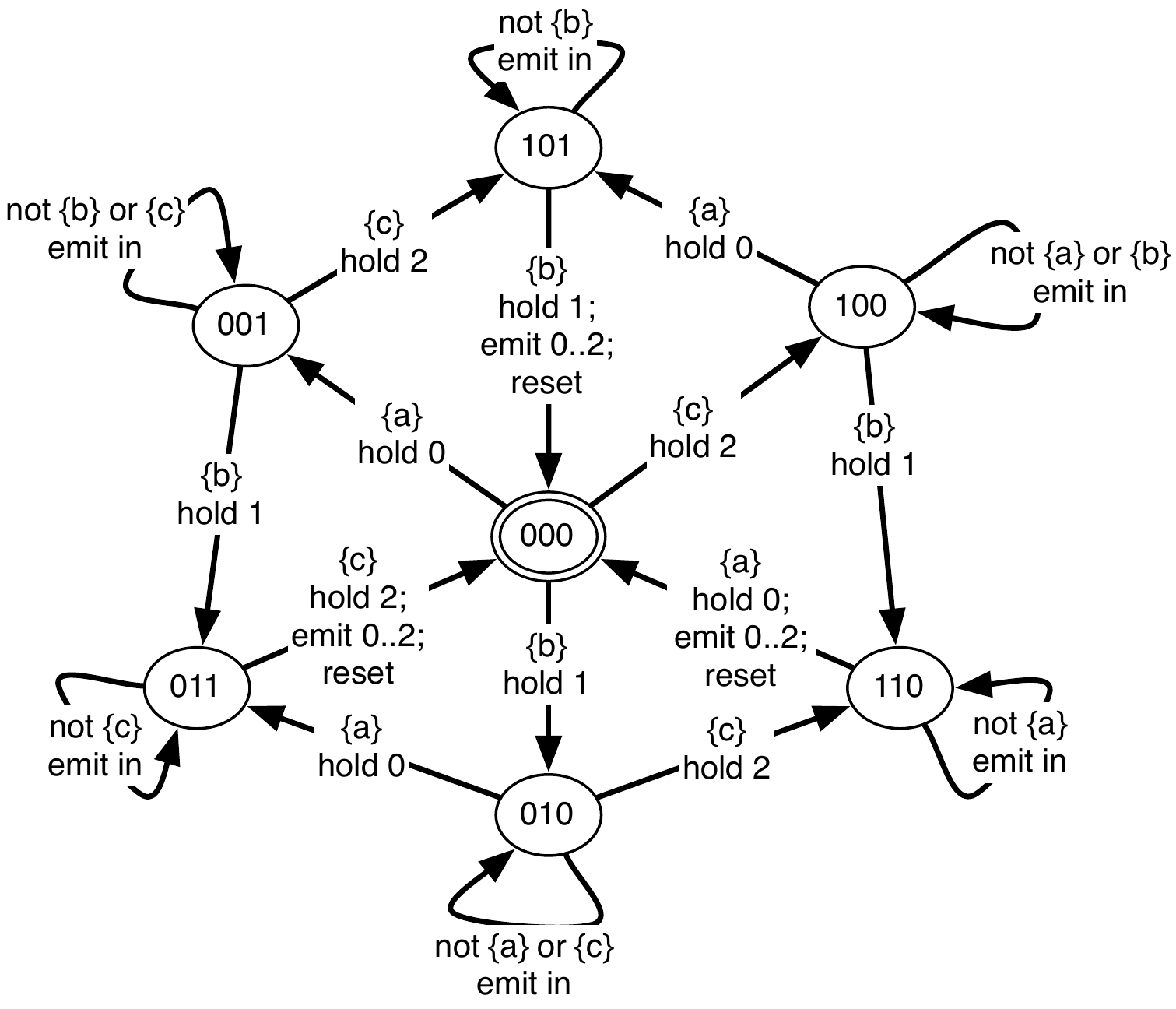}}
\caption{Transducer that merges triplets of $\{a\}$, $\{b\}$, $\{c\}$ records in any order.}\label{fig:trans}
\end{figure}

Finally, it is also possible to express synchronization of multiple
successive records in arbitrary order using label sets
defined as finite-size arrays of bits, as illustrated
by the example in \cref{fig:trans}.  This transducer declares an
array of 3 state bits, \ie 8 possible states.  For every $\mathtt{i}\in\{0,1,2\}$, ``\texttt{\~{}s[i]}'' matches any state of the
label array where bit \texttt{i} is not set; \ie \texttt{\~{}s[0]}
matches labels 000, 010, 100 and 110; \texttt{\~{}s[1]} matches 000, 001,
100 and 101, and \texttt{\~{}s[2]} matches 000, 001, 010 and 011.
``\texttt{t[i]}'' then uses the corresponding guard and
``\texttt{h[i]}'' the corresponding hold variable.  At the end of the
transition ``\texttt{s[i]}'' sets bit \texttt{i} of the bit
array relative to the actual state matched on the left hand side.  In the second transition
specification, \texttt{s[0..2]} applies to the single state where all
bits in the array are set, \ie label 111. It defines an \emph{empty
  transition}, \ie a transition whose action is taken as soon as its origin
state is reached by another transition\footnote{Of course, a cycle of empty transitions is not permitted, otherwise
a transducer could define non-terminating actions.}. Its resulting state specification
unset all bits, \ie resets to label 000. As explained earlier,
unspecified transitions cause the input record to be emitted as-is on the output stream.

This specific example can be reused to synchronize any finite set of record types
by simply extending the size of the arrays; we found it so ubiquitous
in concrete applications that \spnet{} proposes a simplified syntax
for it: \texttt{[| \{a\},\{b\},\{c\} |]}. This extends \snet{}'s ``synchrocell'' concept, as discussed in \cref{sec:syncmotiv}.

\section{Extra-functional specifications}\label{sec:xfun}

\spnet{} also provides \emph{extra-functional combinators}, described
in the following sub-sections, which can be layered on top of
arbitrary networks. In contrast to functional combinators, the
extra-functional combinators were designed so that adding them to a
network does not influence its input-output value relationship.

More specifically, although they \emph{can} influence value
computations, the \spnet{} programmer should not \emph{expect} the
extra-functional combinators to yield the desired functional
effects. This is because any particular implementation of \spnet{} may
not support some of the extra-functional combinators and replace them
with transparent constructs with no effect. For instance, the
``environment awareness'' enables a program to read values that
describe the environment into a record's tag, for example the number
of cores in the resource where a sub-network is currently running. An
application can make use of this number, but acknowledge that a
particular implementation may not support this combinator and always
leave the record tag unmodified instead.

\subsection{Overview}

\spnet{} provides a new primitive network for environment feedback,
noted $\delta$, and the following combinators:
\begin{itemize}
\item replication selection, noted $\mathrm{S}^*$;
\item network labeling, noted $\alpha$, used to designate sub-networks
  in the specification of other extra-functional combinators;
\item environmental exception handling, noted $\beta$;
\item extra-functional requirements on isolation and budget, noted $\theta$ and $\rho$;
\item projection into processing agents, noted $\gamma$ and $\tau$;
\item hardware affinity and mapping, noted $\phi$.
\end{itemize}

\begin{table}
\centering
\begin{tabular}{>{\raggedright}p{.2\textwidth}cp{.15\textwidth}>{\ttfamily}c}
Name & Graphical representation & Algebraic & {\normalfont Source notation} \\
\hline
Replication selection & \vcenteredhbox{\includegraphics[scale=.4]{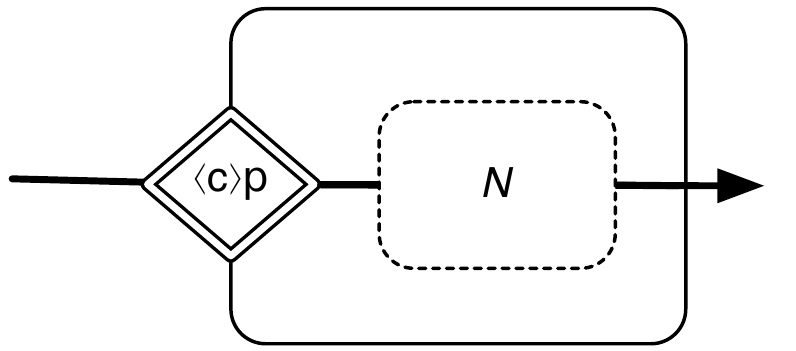}} &
\vcenteredhbox{$\underset{\mathcal{T}^3\times\mathcal{P}\times\mathcal{N}\rightarrow\mathcal{N}}{\mathrm{S}^*_{\langle c\rangle p}(N)}$} & $N$!<$c$>$p$ \\

Labeling & \vcenteredhbox{\includegraphics[scale=.4]{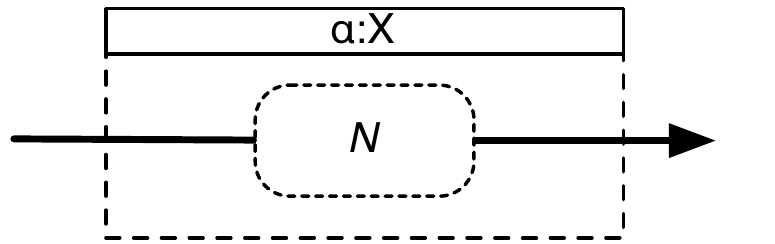}} &
\vcenteredhbox{$\underset{\mathcal{L}\times\mathcal{N}\rightarrow\mathcal{N}}{\alpha_X(N)}$} & $N$'$X$ \\

Environmental exception handling & \vcenteredhbox{\includegraphics[scale=.4]{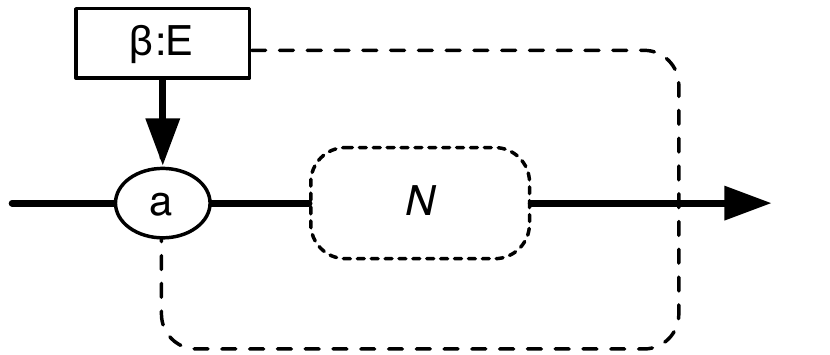}} &
\vcenteredhbox{$\underset{\mathcal{L}\times\mathcal{E}\times\mathcal{F}\times\mathcal{N}\rightarrow\mathcal{N}}{\beta_{X(a=E)}(N)}$} & $N$\$$X$($a$=$E$) \\

Extra-functional isolation & \vcenteredhbox{\includegraphics[scale=.4]{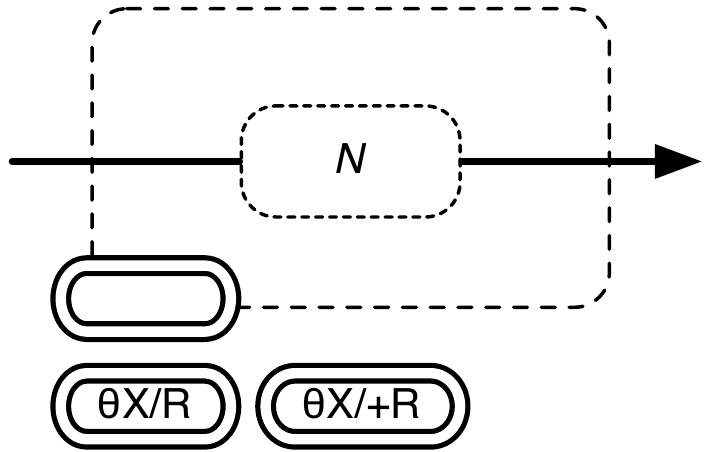}} &
\begingroup\parbox{.15\textwidth}{\centering
$\theta_{X/R}(N)$
$\theta^+_{X/R}(N)$
\tiny $\mathcal{L}\times\mathcal{R}\times\mathcal{N}\rightarrow\mathcal{N}$
}\endgroup
&
\begingroup
\parbox{.15\textwidth}{\centering
$N$/$X$/R
$N$/$X$/+R
}\endgroup \\

Extra-functional budget & \vcenteredhbox{\includegraphics[scale=.4]{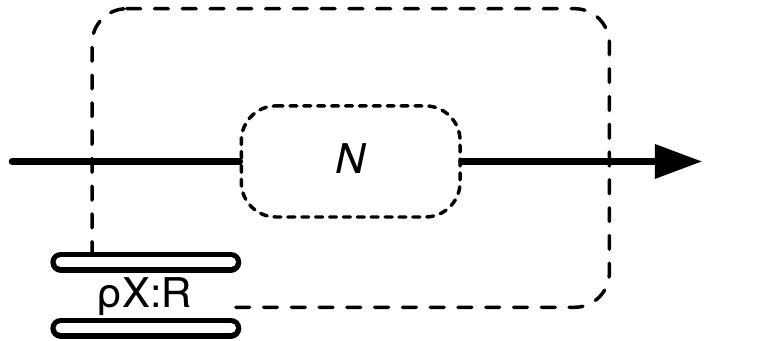}} &
\vcenteredhbox{$\underset{\mathcal{L}\times\mathcal{R}\times\mathcal{N}\rightarrow\mathcal{N}}{\rho_{X:R}(N)}$} & $N$/$X$:$R$ \\

Projections to agents & \vcenteredhbox{\includegraphics[scale=.4]{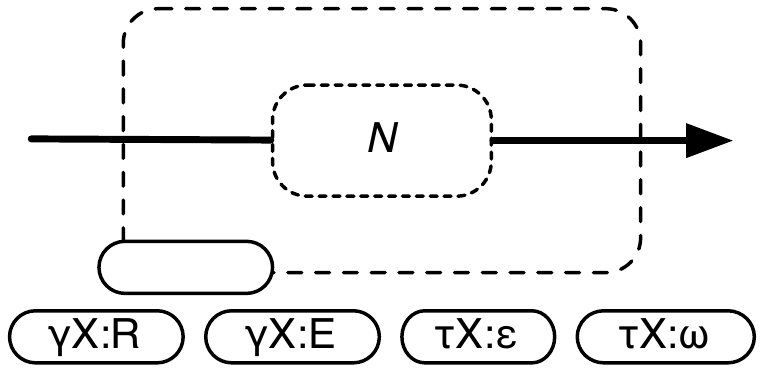}} &
\begingroup\parbox{.15\textwidth}{\centering
$\gamma^\mathrm{e}_X(N)\quad \tau^\omega_X(N)$
$\gamma^\mathrm{r}_X(N)\quad \tau^\epsilon_X(N)$
\tiny $\mathcal{L}\times\mathcal{N}\rightarrow\mathcal{N}$
}\endgroup
&
\begingroup
\parbox{.15\textwidth}{\centering
$N$/$X$!ge
$N$/$X$!gr
$N$/$X$!to
$N$/$X$!te}\endgroup \\

Resource affinity and assignment & \vcenteredhbox{\includegraphics[scale=.4]{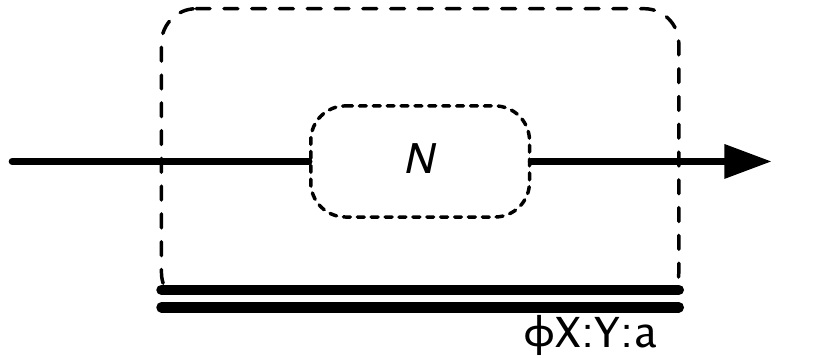}} &
$\underset{\mathcal{L}^2\times\mathcal{P}\times\mathcal{N}\rightarrow\mathcal{N}}{\phi_{X@Y:a}(N)}$ & \texttt{$N$/$X$@$Y$:$a$} \\

Environment awareness & \vcenteredhbox{\includegraphics[scale=.4]{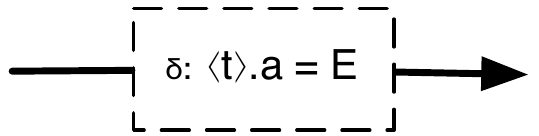}} &
\vcenteredhbox{$\underset{\mathcal{T}\times\mathcal{F}\times\mathcal{D}\rightarrow\mathcal{N}}{\delta_{\langle t\rangle.a}(E)}$} & [<$t$>.$a$=$E$] \\

\end{tabular}
\caption{Notations for \spnet{}'s extra-functional primitive network and combinators.}\label{tab:efequiv}
\end{table}

The corresponding notations are given in \cref{tab:efequiv}.
All these constructs, except for $\beta$, $\delta$ and $\mathrm{S}^*$, are
functionally neutral: they do not change the input-output relationship
of the encapsulated network.

Both $\beta$ and $\delta$ read and modify scalar values in records,
and thus establish a bridge between functional and extra-functional
semantics; $\mathrm{S}^*$ explicitly manipulates network
replicas and thus can influence the functional semantics via output
reordering and state management.  However, we
consider that an \spnet{} specification is only well-formed if its
input-output relationship stays valid for the application when all
extra-functional combinators or $\delta$ are elided.

\subsection{Replication selection}\label{sec:ersel}

The network combinator $\mathrm{S}^*$ expresses replication selection
with optional extra-functional choice between replicas. This construct
can be used to manually fine-tune the exploitation of parallel
hardware; it is not intended for use during an initial specification
or to when determining functional correctness of a specification.

The construct $S^*_{\langle c\rangle p}(N)$ is a network specification if $N$ is a network
specification. It is also noted ``\texttt{N!<c>p}'' in source
form. $N$ is called the \emph{inner network}; $\langle c\rangle$ is
the \emph{selection tag} and $p$ is called the \emph{selection policy}. 

The semantics are as follows:
\begin{itemize}
\item the network $\mathrm{S}^*_{\langle c\rangle p}(N)$
  maintains an ordered \emph{processing set} of indexed replicas of
  $N$ over time;
\item whenever it receives a record tagged by $\langle c\rangle$:
\begin{itemize}
\item if the tag's scalar value is stricly positive, it forwards the
  record to the replica indexed by the tag's scalar value, creating
  the replica if necessary;
\item if the scalar is negative, it forwards the record to the
  replica indexed by the absolute value, then signals termination to the replica;
\end{itemize}
\item if $\langle c\rangle$'s value is zero, and for all records of another type,
  it routes the record to a replica
  according to the policy (described below).  Moreover, if $\langle c\rangle$'s value is zero,
  then the value of $\langle c\rangle$ is automatically set upon entry by $\mathrm{S}^*$
  to the index of the selected replica.
\end{itemize}

If there are no replicas currently defined upon receiving a $\langle
c\rangle$ record where $\langle c\rangle$'s value is zero or upon
receiving a record of another type, or if $\langle c\rangle$'s value
is negative and there is currently no replica indexed by the absolute
value, the record is forwarded directly to the output stream.

The following policies are available:
\begin{itemize}
\item $\mathrm{S}^*_\mathsf{e}$ for \emph{even} distribution: best effort is made
  to distribute the records evenly to the current replicas;
\item $\mathrm{S}^*_\mathsf{lr}$ for \emph{last replica}: records are distributed
  to the last replica selected by $\langle c\rangle$;
\item $\mathrm{S}^*_{\mathsf{la}, \mathsf{ha}}$ for \emph{lowest or highest
  available replica}: records are distributed to any replica currently
  able to accept input, preferring replicas with the lowest index or
  highest index respectively.
\end{itemize}

If the policy is not specified, it defaults to the even distribution.

Activity and termination are further determined for $\mathrm{S}^*$ like for
selection and replication composition earlier.

\subsection{Identifiers for run-time activities}\label{sec:rtident}

As discussed in \cref{sec:fun} and more specifically in \cref{sec:liveness},
any static network specification translates, at run-time, into
zero or more replicas,
and for each replica, into a lifecycles of zero or more activations
  before eventual termination.
Any run-time activity, either communication or processing or records,
can thus be traced back to the static specification using:
\begin{itemize}
\item for transformations, the path to the sub-network defining the
  transformation, augmented at each level of nesting with an
  identifier for the replica where the transformation takes place and
  an identifier for the activation of that replica;
\item for records, the identity of the transformation that
  has produced the record, augmented with the causal index\footnotemark{} of that record.
\end{itemize}
\footnotetext{The causal index is due to multiplicity: any input record processed by
a box can cause zero, one or more output records, and these can need
to be distinguished by other means than their type.}
  
To identify run-time activities, \spnet{} standardizes the notion
of \emph{network index}.
A network index is a list of triplets $(x,y,z)$ where $x$ is the functional
path through the specification, $y$ an identifier for the replica and
$z$ an identifier for a particular activation.  Functional paths and
replica identifiers are defined ``naturally'' for functional
combinators:
\begin{itemize}
\item for simple composition $\mathrm{C}$ and selection $\mathrm{S}$,
  the functional path designates the position of the sub-network in
  specification order, and the replica identifier is typically 0 (although
  that may be changed by extra-functional combinators below);
\item for ordered replicated composition $\mathrm{C}^*$, the functional
  path is the position of the replica in the dynamic
  unfolding\footnotemark{}, and the replica identifier is merely
  unique for that position;
\item for unordered replicated composition $\mathrm{C}^!$, the functional
  path is an identifier for the processing subsequence on the input,
  the replica identifier is merely unique for that subsequence;
\item for replicated selection $\mathrm{S}^*$, the functional path is the
  scalar value that identifies which alternative to use, and the
  replica identifier is typically equal to the static path.
\end{itemize}

\footnotetext{The position is given prior to garbage collection, \ie
the index may be larger than the dynamic depth.}

For example, consider the network 
\[
\mathrm{C}\left(N, \mathrm{S}\left(\mathrm{C}^*\circ\mathrm{C}\left(M, O\right), P\right)\right)
\]
or ``\texttt{N..((M..O)*|P)}'' in source form.  An activity on behalf
of the 3rd activation of the 12th replica of network $N$ would be
identified by $[(0,11,2)]$.  The value 0 indicates the first argument
of the first $\mathrm{C}$ combinator. Likewise, the index
$[(1,0,12);(0,0,12);(12,0,12);(0,0,12)]$ identifies an activity on
behalf of the second position of the outer $\mathrm{C}$, of the first
position of $\mathrm{S}$, of the 13th unfolding of $\mathrm{C}^*$ and
at the first position of the inner $\mathrm{C}$, \ie in $M$.

From network indices we derive the notion \emph{functional network
  indices}, which are lists formed by taking the first index of each
element in a full network index. This notion is equivalent
to the network indices defined in~\cite[App.~B.3]{holzenspies.10}.
Functional network indices are often sufficient to establish
functional causality for observed values, but they may lose information
about extra-functional causes.

\subsection{Network labeling and selection}\label{sec:netlabel}

The $\alpha$ combinator helps naming and
identifying run-time instances of subnetworks.  The
construct $\alpha_X(N)$ (``\texttt{N'X}'' in source) is a network specification if $N$ is a network
specification.
$X$ is
called the \emph{network label}.  The functional semantics of $\alpha_X(N)$ are those of
$N$.

\begin{figure}
\centering
\includegraphics[scale=.4]{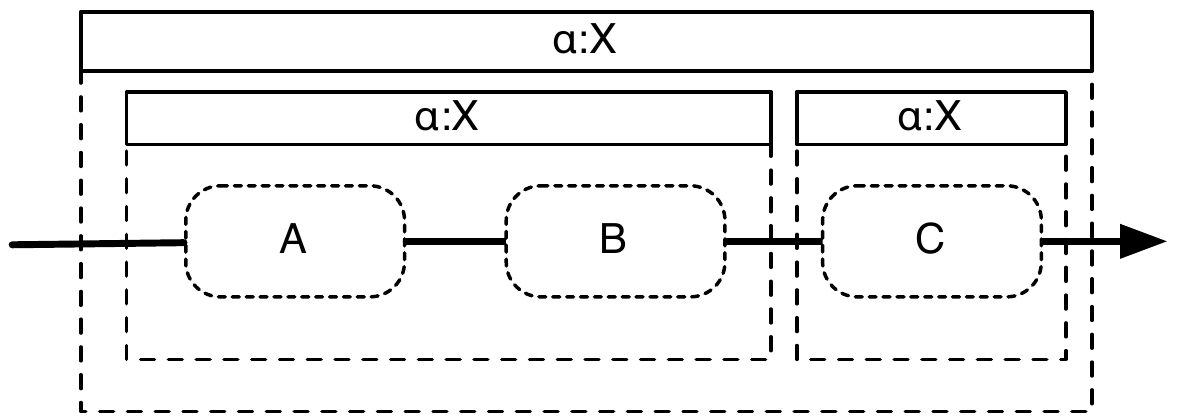}

``\texttt{ ((A..B)'X..C'X)'X }''
\caption{Graphical representation of $\alpha_X\circ\mathrm{C}(\alpha_X\circ\mathrm{C}(A,B), \alpha_X\circ C)$}
\label{fig:alphaex1}
\end{figure}

\begin{figure}
\centering
\includegraphics[scale=.4]{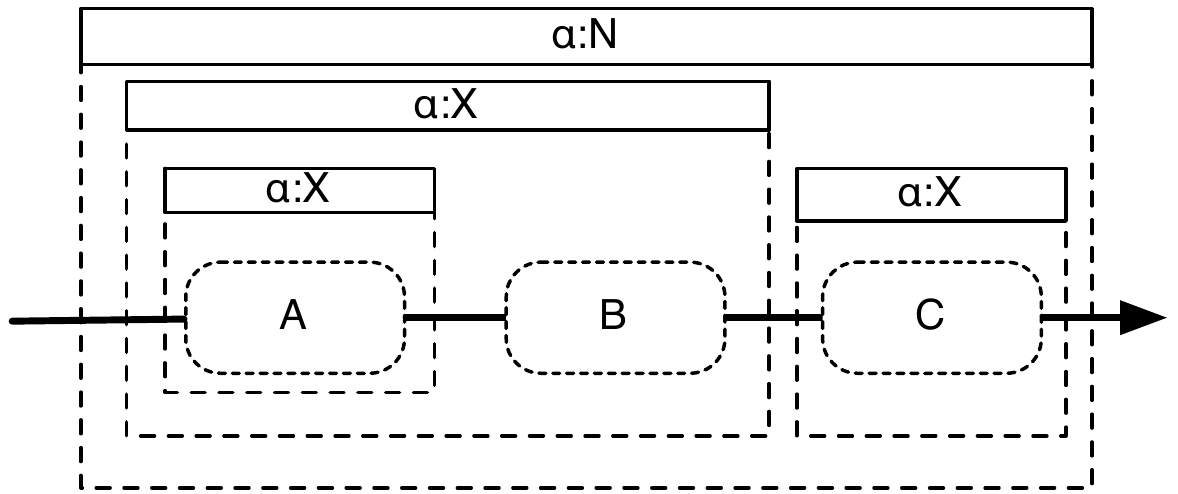}

``\texttt{ ((A'X..B)'X..C'X)'N }''
\caption{Graphical representation of $\alpha_N\circ\mathrm{C}(\alpha_X\circ\mathrm{C}(\alpha_X\circ A,B), \alpha_X\circ C)$}
\label{fig:alphaex2}
\end{figure}

Network labels are used as anchors by the extra-functional features
defined hereafter. There are two possible uses of anchors:
\begin{itemize}
\item the constructs $\beta$, $\theta$, $\rho$, $\gamma$, $\tau$ and $\phi$ refer to the set of
  \emph{outermost inner network(s)} with some label. For example in
  \cref{fig:alphaex2}, $N$'s outermost inner networks labeled by $X$
  are $\mathrm{C}(A,B)$ and $C$;
\item the constructs $\delta$ and $\phi$ refer to the \emph{innermost outer network} using some label. For example
  in \cref{fig:alphaex1}, $A$'s innermost outer network labeled by $X$ is $\mathrm{C}(A,B)$,
 whereas $C$'s is itself.
\end{itemize}

Note that \spnet{}'s construct ``\texttt{net $X$ ...}'' in source
notation also implicitly generates a use of $\alpha_X$ around the enclosed
network specification in \spnet{}.

\subsection{Environmental exception handling}

The network combinator $\beta$ provides a facility to handle
extra-functional exceptions in the run-time environment.  The
construct $\beta_{X(a=E)}(N)$ (``\texttt{N\$X(a=E)}'' in source) is a network specification if $N$ is a
network specification. $E$ is called the
\emph{exception specification}; $a$ is called the \emph{exception
  label}. $X$, if specified,
is called the \emph{target network selector}. 

$\beta$ applies to all the outermost inner networks labelled by $X$,
or to the inner network directly if $X$ is omitted. The selected
network(s) are called \emph{target networks}.
Semantically, the input of $\beta_{(a=E)}(N)$ is provided to $N$
in-order, unchanged.  If the processing of an input record $r$ by any
of the target networks causes an uncaught extra-functional exception of type
$E$, it is reported as
follows:
\begin{itemize}
\item all outputs of $N$ produced causally from $r$ are not output from $\beta(N)$;
\item if $r$ has no field labelled $a$, or if $a$'s value is non-zero,
  then the exception is propagated to the enclosing network;
  otherwise:
\begin{itemize}
\item the state of $N$ is restored to prior to $r$'s input; and
\item $r$ is re-injected as input to $N$ with
  its scalar field $a$ modified to a non-zero value (indicating an exception has
  occurred).
\end{itemize}
\item whenever an exception escapes the scope of a $\beta$ network,
  all the replicas of the inner network are terminated, even if
  they are active. Subsequent input records cause a new instantiation.
\end{itemize}

A $\beta$ network may let subsequent records of the input flow in the
protected network before the faulty record is retried without
violating causality. This is possible to implement if the protected
network is stateless, if the environment supports transactions and an
aborted transaction has no impact on the processing of subsequent
records, or if the specific application tolerates such reordering. To
obtain determinism, that is, preservation of input order,
$\mathrm{R}\circ\beta$ can be used.

Any exceptions generated by sub-networks others than the target
network(s) are not caught and propagated to the enclosing environment instead.

\subsection{Extra-functional isolation}\label{sec:efreqs}

The network combinators $\theta$ and $\theta'$ express
extra-functional isolation requirements on the execution
environment. The constructs $\theta_{X/R}(N)$ and $\theta^+_{X/R}(N)$
(``\texttt{N/X/R}'' and ``\texttt{N/X/+R}'' in source) are network
specifications if $N$ is a network specification. $R$ is called the
\emph{isolation property}; $X$, if specified, forms the target network
selector. Both $\theta(N)$'s and $\theta^+(N)$'s functional semantics
are those of $N$. Like $\gamma$ and $\tau$, $\theta$ and $\theta^+$
apply to the outermost inner networks labelled by $X$, or to the inner
network directly if $X$ is omitted; the selection is also called
target network(s).
The functional semantics of $\theta(N)$ and $\theta^+(N)$ and are those of $N$.

Extra-functionally, 
$\theta_{R}$ specifies that the execution
environment guarantees that all replicas of the target networks are
isolated \emph{from each other} relative to property $R$. In contrast,
$\theta^+_R$ specifies that the target replicas are isolated \emph{from
  each other and also from their enclosing network} relative to $R$. In particular
with $\theta$, the management activities of the enclosing network need not be isolated
from the replicas of the target networks; only $\theta^+$ guarantees this isolation.

\snet{} proposes the following isolation specifications:
\begin{itemize}
\item $\theta_{/\mathsf{f}}$ for \emph{relative progress independence}
  (fairness), \ie the progress of each replica not starved on input or
  blocked on output is guaranteed independently from other replicas
  (and from the enclosing network with $\theta^+$). This may be
  implemented using preemptive time sharing;
\item $\theta_{/\mathsf{b}}$ for \emph{relative bandwidth independence}, \ie the
  internal bandwidth of processors and channels onto which each
  replica is mapped is reserved and free of contention from other
  replicas (and from the enclosing network with $\theta^+$). This may
  be implemented using round-robin real-time scheduling on one
  processor/channel, or via true hardware parallelism;
\item $\theta_{/\mathsf{s}}$ for \emph{relative storage independence}, \ie 
  the storage allocated by the running components and network management is sourced from 
  storage pools separate between replicas (and from the enclosing network with $\theta^+$);
\item $\theta_{/\mathsf{p}}$ for \emph{relative energy supply independence}, \ie
  the power demands of each replica are satisfied independently from
  the power usage of other replicas (and from the enclosing network with $\theta^+$).
\end{itemize}

An implementation of \spnet{} may be unable to satisfy a $\theta$ requirement
at run-time. In this case, best effort is made to report this inability
statically; otherwise an exception of type $\mathsf{Violation}(R)$ is
raised at the first network activation without the required guarantee.

\subsection{Extra-functional budget}\label{sec:efbudget}

The network combinator $\rho$ expresses extra-functional budget
requirements on the execution environment. The construct
$\rho_{X:R}(N)$ (``\texttt{N/X:R}'' in source) is a network
specification if $N$ is a network specification. $R$ is called the
\emph{budget specification}; $X$, if specified, forms the target
network selector. $\rho(N)$'s functional semantics are those of
$N$. Like $\gamma$, $\tau$ and $\theta$ previously, $\rho$ applies to
the outermost inner networks labelled by $X$, or to the inner network
directly if $X$ is omitted; the selection is also called target
network(s). The functional semantics of $\rho(N)$ are those of $N$.

Extra-functionally, $\rho_{X:R}(N)$ specifies that the execution
environment caps the extra-functional budget $R$ available to all replicas of
the target networks, to a proportion of the budget available
to the surrounding network. The following budget specifications are
supported:
\begin{itemize}
\item $\mathsf{mp}(x)$ for \emph{maximum power}, \ie the
  amount of power collectively consumed by all target networks;
\item $\mathsf{mc}(x)$ for \emph{maximum storage},
  \ie the amount of storage collectively allocated;
\item $\mathsf{mfl}(x)$ and $\mathsf{mll}(l)$ for
  \emph{maximum first/last latency}, \ie the maximum duration between
  the moment a record is input by a replica and the first/last output
  causally produced;
\item $\mathsf{mti}(x)$ and $\mathsf{mto}(x)$
  for \emph{maximum input/output throughput}, respectively;
\item $\mathsf{mdla}(x)$, $\mathsf{mdaa}(x)$ and $\mathsf{mdpa}(x)$
  for \emph{maximum dynamic liveness/activity/agent arity}, respectively.
\end{itemize}

The parameter $x$ establishes a budget relative to the budget
available to the enclosing network. This can be either a ratio (\eg
10\%) or maximum absolute value (\eg 10ms). The coordination layer
makes a best effort to enforce the requirement, possibly throttling
the execution strategy and processing rates to match the
constraint. If a maximum value cannot be satisfied from the outer
budget, or when a network is known to behave in violation with
the requirement, an exception of type $\mathsf{Violation}(R)$ is
raised at run-time.  This may happen while a box is computing; for
example, with $\rho_\mathsf{mfl}$ if a latency bound is exceeded while
a box is still transforming its input record, the behavior can be
aborted preemptively.

\subsection{Projections: mapping specifications into processing agents}\label{sec:projs}

The network combinators $\gamma$ and $\tau$ express how run-time
agents are spawned to process activations. $\gamma$ chooses between
entity-centric and record-centric projections, and thus helps control
locality and the trade-off between computation and communication (\cf \cref{sec:projmotiv}).
$\tau$ decides the lifetime of agents, and thus helps control the
trade-off between jitter and throughput (\cf \cref{sec:taumotiv}).

The constructs $\gamma^\mathrm{e}_X(N)$, $\gamma^\mathrm{r}_X(N)$,
$\tau^\omega_X(N)$ and $\tau^\epsilon_X(N)$ are network
specifications if $N$ is a network specification. They are noted
\texttt{N/X!ge}, \texttt{N/X!gr}, \texttt{N/X!to}, \texttt{N/X!te} in
source form. $X$, if specified,
is the target network selector. Their functional semantics
are those of $N$.  As with the previous combinators, their semantics apply to the
target network(s) selected by $X$, or to the inner network
directly if $X$ is omitted. 

\begin{itemize}
\item $\gamma^\mathrm{e}$ specifies that the target networks should be mappped
  to run-time agents using an \emph{entity-centric projection}, that is,
  agents are created for the entities in the specification. For
  example, for the network $\mathrm{C}(A,B)$ one agent is created for $A$ and
  another for $B$. Agents communicate over buffered channels, which implement
  the \spnet{} streams, and terminate when the corresponding functional
  entity terminates as per \cref{sec:liveness};
\item $\gamma^\mathrm{r}$ specifies that the target networks
  should be mapped to run-time agents using a \emph{record-centric
    projection}, that is, agents are created for each successive input
  records. For example, for $\mathrm{C}(A,B)$ one agent for a first
  input record $r_1$ executes code for $A(r_1)$ then code for
  $B(A(r_1))$; another agent for $r_2$ executes code for $A(r_2)$ then
  code for $B(A(r_2))$;
\item $\tau^\omega$ specifies that agents created for the target networks
  should be \emph{lingering}: once an agent is created, it tries to
  consume records/entities as long as it can. Agents with
  $\gamma^\mathrm{r}$ may terminate at transducers and reordering points,
  at the latest at the output edge of $N$. 
  Agents with $\gamma^\mathrm{e}$ terminate when a network terminates;
\item $\tau^\epsilon$ specifies that agents for the target networks should be
  \emph{ephemeral}: an agent only runs for the duration of an
  ``elementary'' processing: with $\gamma^\mathrm{e}$, only for one agent
   cycle (\cf \cref{sec:liveness}); with $\gamma^\mathrm{r}$, only for one step through
  the network graph.
\end{itemize}

Agents correspond to cooperatively scheduled tasks, threads, processes
or virtual machines depending on the implementation and the
requirements expressed via $\theta$ and $\rho$ (\cf \cref{sec:efreqs,sec:efbudget}).  Agents are identified by the full
network index of the activation that created them. Agents may run
concurrently, either interleaved over time on sequential processors or
simultaneously on parallel hardware. With $\gamma^\mathrm{e}$ the concurrency
of agents stems from the concurrency between the logical entities in
the specification; whereas with $\gamma^\mathrm{r}$ the agent concurrency stems
from stream concurrency.

From the projection of specifications into agents we derive
the notion of \emph{dynamic agent arity} for a sub-network,
which is the current number of running agents for this sub-network.

\subsubsection{Composability of $\gamma^\mathrm{r}$ and $\gamma^\mathrm{e}$}

Any $\gamma^\mathrm{r}$ sub-network maps to a set of agents under
control of a single ``input'' agent which reads the input records
of the entire sub-network from its surrounding environment. Moreover, all the
agents of a $\gamma^\mathrm{r}$ network \emph{eventually} synchronize so
that all final outputs of the network appear as if produced by a
single ``output'' agent. Therefore, any $\gamma^\mathrm{r}$ network is a
valid operand for combinators captured within a $\gamma^\mathrm{e}$
combinator. For example, the network 
\begin{center}
$\gamma^\mathrm{e}\circ\mathrm{S}\left(\gamma^\mathrm{r}\circ\mathrm{C}(M,N),\gamma^\mathrm{r}\circ\mathrm{C}(O,P)\right)$

``\texttt{ ( (M..N)/!gr | (O..P)/!gr ) /!ge }''

(also: ``\texttt{ ( (M..N)'x | (O..P)'x ) /!ge /x!gr }'')
\end{center}
expresses that the transformations for $\mathrm{C}(M,N)$ and
$\mathrm{C}(O,P)$ should be carried out in a record-centric fashion,
but that different (groups of) entity-centric agents should be
used to create a pipeline-like parallelism between the subnetworks of
the selection.

Conversely, all agents created for a $\gamma^\mathrm{e}$
network can be drained of input, and their left-over state serialized
to be re-instantiated later or elsewhere. Therefore, any $\gamma^\mathrm{e}$
network is also a valid operand for a
$\gamma^\mathrm{r}$ network. 

\subsubsection{Sequential execution}

$\rho_\mathsf{mdpa}$ can interact with $\gamma$ to obtain
sequential execution.  For example,
$\gamma^\mathrm{r}\circ\rho_{\mathsf{mdpa}(1)}$ forces each
input record to be processed, in turn, sequentially through the inner
network.  $\gamma^\mathrm{e}\circ\rho_{\mathsf{mdpa}(1)}$ forces
the first entity to process the entire input stream sequentially,
accumulating all its output records in temporary buffers, only then
lets the second entity process all the intermediate records
sequentially, and so forth.

\subsection{Hardware affinity and mapping}\label{sec:phi}

Resource mapping assumes that hardware resources are structured in a
tree that reflects locality and granularity, and where the leaves are
processing and storage units; for example cluster nodes at the top
level, then processor sockets, then processor cores, with hardware
threads and memories as leaves.  The tree need not be homogeneous nor
balanced. Each node in the tree is \emph{enumerable} (by index from
its parent node) and some nodes may also be annotated by
\emph{metadata} that indicate special features, \eg hardware
accelerators or specialized (IP) cores.

There are two parts to resource mapping in \spnet{}: \emph{assignment}
and \emph{placement}.  Assignment establishes a mapping from a set of
sub-networks sharing a common network label to a node in the hardware
resource tree, possibly higher than leaves. Placement occurs at the
point processing agents are created, using the resources
previously assigned to the corresponding sub-network. Once placed, an
agent stays placed at the same resources until it terminates;
the $\tau$ introduced earlier in \cref{sec:projs} thus helps control opportunities for work migration.

The combinator $\phi$ only helps manage assignment of hardware
resources. Placement is automatically managed by the coordination
layer to satisfy the demands of $\theta$ and $\rho$ within the
resources assigned by $\phi$. However, $\delta(\mathsf{h})$ (\cf
\cref{sec:envaware}) can observe placement after it is computed.

\subsubsection*{Semantics in \spnet{}}

The construct $\phi_{X,Y:a}(N)$ (``\texttt{N/X@Y:a}'' in source) is a
network specification if $N$ is a network specification. $a$ is called
the \emph{assignment specification}; $X$ is the target network
selector and $Y$ is called \emph{origin selector}.  The functional
semantics of $\phi_{X,Y:a:p}(N)$ are those of $N$.

\begin{figure}
\centering
\includegraphics[scale=.4]{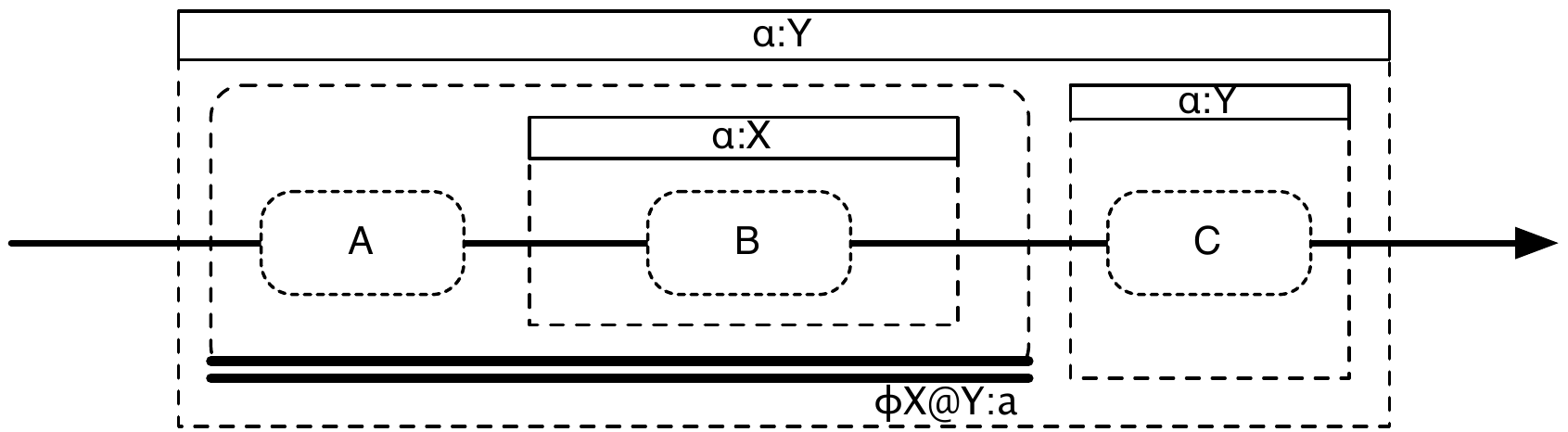}

``\texttt{ ( (A..B'X) /X@Y:a ..C'Y )'Y }''
\caption{Graphical representation of $\alpha_Y\circ\mathrm{C}(\phi_{X,Y:a}\circ\mathrm{C}(A,\alpha_X\circ B), \alpha_Y\circ C)$}
\label{fig:phiex1}
\end{figure}

\begin{figure}
\centering
\includegraphics[scale=.4]{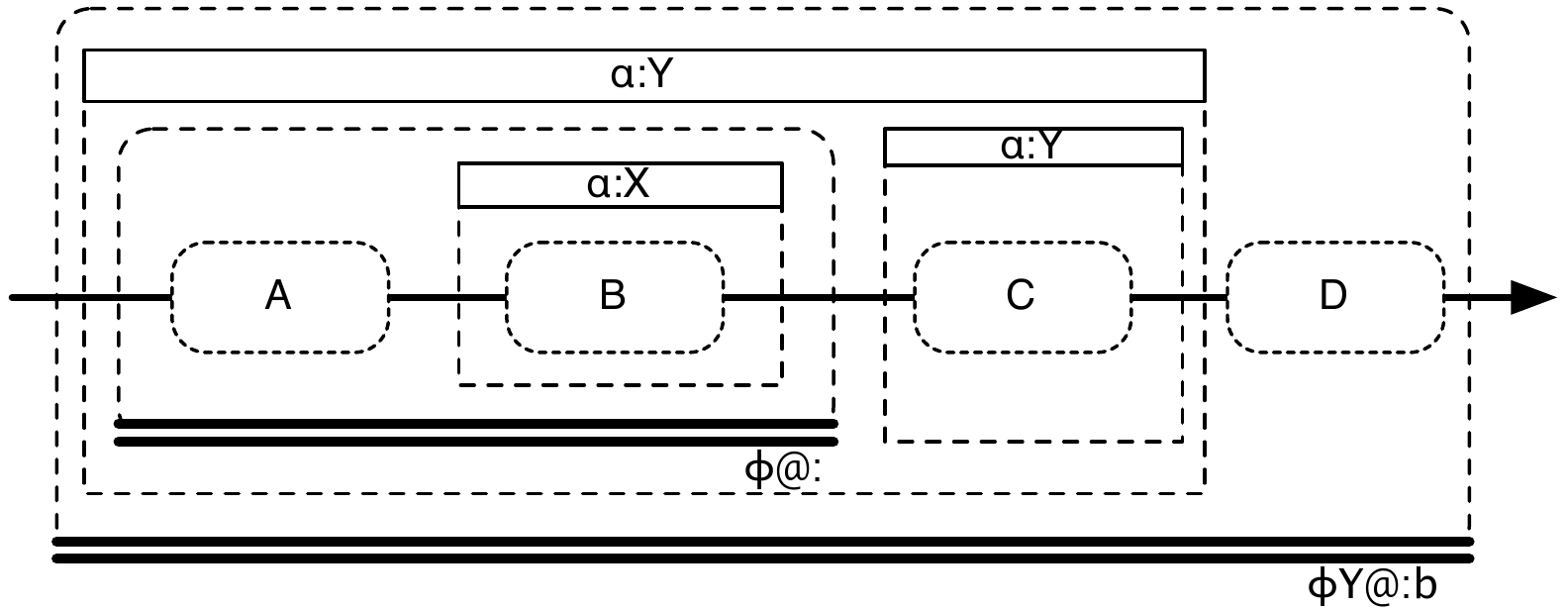}

``\texttt{ ( ( (A..B'X) /@ ..C'Y )'Y .. D) /Y@:b }''
\caption{Graphical representation of $\phi_{Y,:b}\circ\mathrm{C}(\alpha_Y\circ\mathrm{C}(\phi\circ\mathrm{C}(A,\alpha_X\circ B), \alpha_Y\circ C), D)$}
\label{fig:phiex2}
\end{figure}

Like the previous combinators, $\phi$ applies to the
target network(s) selected by $X$ or the inner network direcly if $X$
is omitted.  It also refers to the innermost outer network labelled by
$Y$, or to the innermost outer network targeted by an outer $\phi$ if
$Y$ is omitted; this selection is called \emph{origin network}. For
example in \cref{fig:phiex1}, $\phi$'s target network is $B$ whereas
its origin network is the entire group $\mathrm{C}(A,B,C)$. In
\cref{fig:phiex2}, the inner $\phi$'s target network is
$\mathrm{C}(A,B)$ because there is no target selector. Its origin
network is the one labeled by the outer $Y$, \ie the entire group
$\mathrm{C}(A,B,C)$, because that is the innermost outer network
targeted by the outer $\phi$.

Extra-functionally, $\phi$ assigns sub-resources from the origin
network to replicas of the target networks. If $a$ is omitted, the
construct inherits the resources assigned to the origin network. For
example in \cref{fig:phiex2}, the resources assigned
are inherited from the outer $Y$ (therefore this inner $\phi$
does not specify anything useful).

The following assignment
specifications are supported:
\begin{itemize}
\item $\mathsf{share}(x)$: all replicas of the target networks
  are commonly assigned the sub-resources selected by $x$;
\item $\mathsf{split}(x)$: each new replica
  is assigned one of the sub-resources selected by $x$.
\end{itemize}

The syntax of $x$ will be detailed in future work; it permits either
to select a path in the resource tree (to achieve resource
partitioning) or filter sub-trees based on a predicate on metadata (to
select specialized resources), relative to the resources assigned at the
origin network. If there are no sub-resources selected, or if new
replicas run out of unassigned sub-resources, an exception of type
$\mathsf{Exhaustion}$ is raised. $\delta(\mathsf{h})$ can be used to inspect the arity of the
selected sub-resources and the number of sub-resources yet unassigned.

\subsection{Environmental awareness}\label{sec:envaware}

Environment observations are noted $\delta_{\langle t\rangle.a}(E)$ in
algebraic form or ``\texttt{[<t>.a=E]}'' in source form.  $\langle t\rangle$ is
called the \emph{matching tag}; $a$ is called the \emph{payload
  field}, and $E$ is the \emph{environment function}.

A $\delta$ network reproduces its input records to its output stream
unchanged and in-order, except that for each record consumed that
matches tag $\langle t \rangle$, the value of the field $a$, if
present, is updated depending on $E$ before forwarding the record.  If
$\langle t\rangle$ is not specified, $a$ is modified in all records
that contain the field label, regardless of tag. If $a$ is not
specified, the scalar value of the tag itself is modified.
Implementations of \spnet{} provide at least the following environment
functions:

\begin{itemize}
\item $\mathsf{time}(g)$: difference between
  the current time and the scalar's previous value;
\item $\mathsf{c}(X,i,g)$: current amount of storage used by
  activities;
\item $\mathsf{p}(X,i,g)$: 
  current power used by activities;
\item $\mathsf{ti}(X,i,g)$, $\mathrm{to}(X,i,g)$: 
  current input/output throughput;
\item $\mathsf{fl}(X,i,g)$, $\mathsf{ll}(X,i,g)$: 
   current first/last output latency (from the input of a
  record $r$ and the first/last output causally produced from
  $r$);
\item $\mathsf{dla}(X,i)$, $\mathsf{daa}(X,i)$ and $\mathsf{dpa}(X,i)$: current
  dynamic liveness/activity/agent arity, respectively;
\item $\mathsf{h}(X,i,p)$: value of the
  property $p$ of the hardware resources currently assigned.
\end{itemize}

\begin{figure}
\centering
\includegraphics[scale=.4]{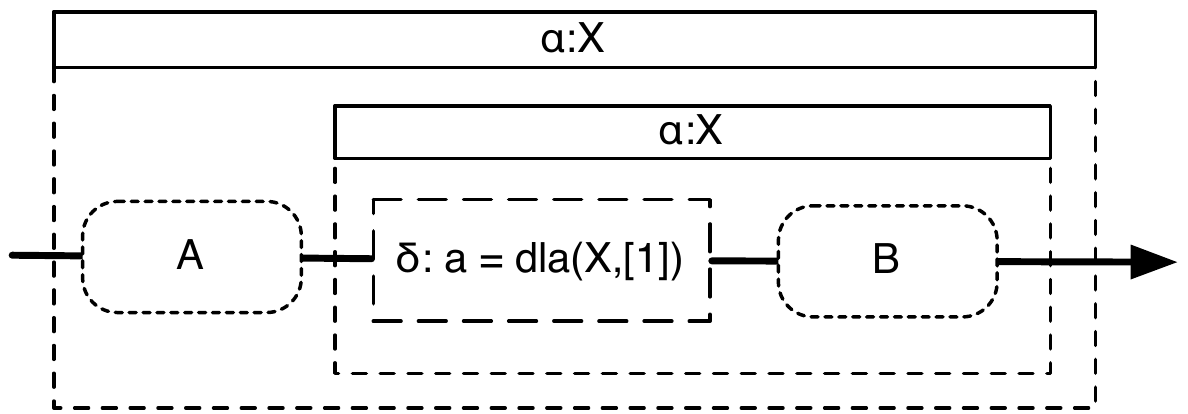}

``\texttt{ ( A..([a=dla(X,[1])]..B)'X )'X }''
\caption{Graphical representation of $\alpha_X\circ\mathrm{C}(A, \alpha_X\circ\mathrm{C}(\delta_a(\mathsf{dla}(X,[1])),B))$}
\label{fig:deltaex1}
\end{figure}

For each function parameterized
by a network label $X$ and functional index $i$, the observation
pertains to the sub-network identified by $i$ relative to the
innermost outer network labeled by $X$. For example in \cref{fig:deltaex1},
the observation pertains to network $B$. The network index is necessary
to disambiguate which network to observe when the same label is used
multiple times; future work will explore semantics
to gather observations across multiple sub-networks.

Throughputs are expressed in records per $g^{-1}$ seconds, power in
$g^{-1}$ watts, time and latencies in multiples of $g^{-1}$ seconds,
and storage in multiples of $g$ bytes. The values are averaged over
time using a sliding window or exponential smoothing.  If the
environment does not support a function $E$ at run-time, an exception
of type $\mathsf{Unimplemented}$ is raised.

\subsection{Implementation services}\label{sec:implsvc}

Next to the application specifications constructs defined so far,
the \spnet{} executionb environment provides the following services, available
to the operator of a running application:

\begin{itemize}
\item
\emph{lookup}, which
given a run-time event and a network label, produces the network
index \emph{relative} to the most inner
$\alpha$ construct that identifies the run-time event.

\begin{figure}
\centering
\includegraphics[scale=.4]{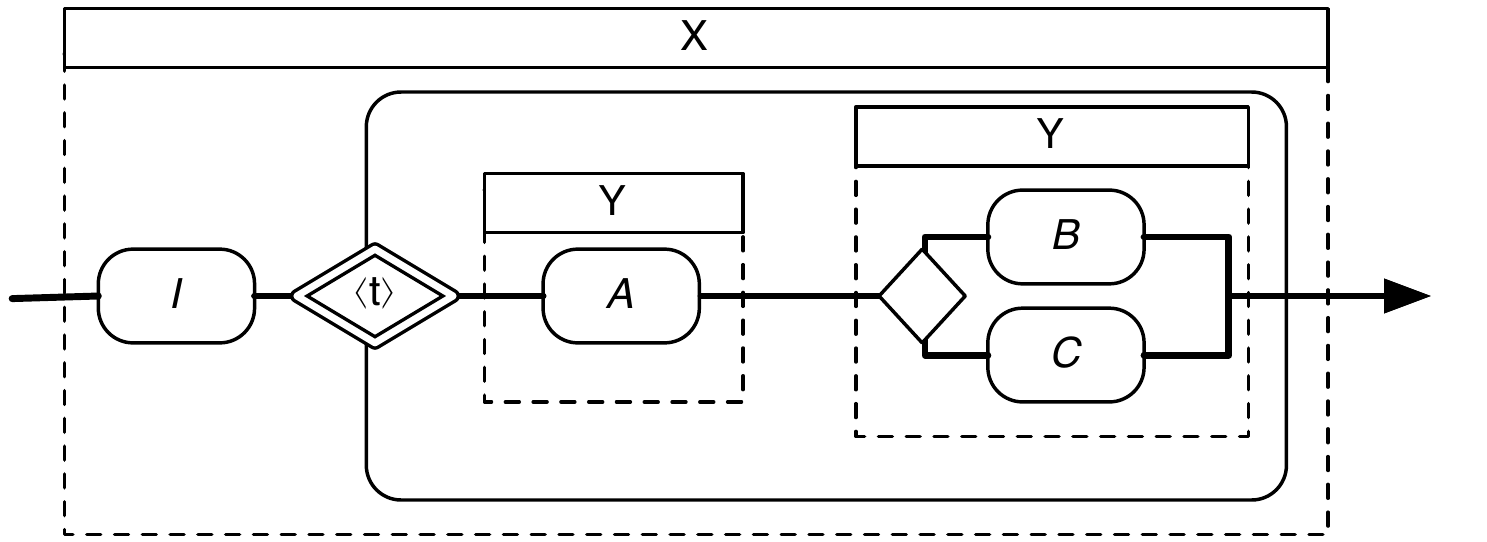}

\scriptsize 
``\texttt{($I$..($A$'Y..($B$|$C$)'Y)!<t>)'X}''
\caption{Graphical representation of
$\alpha_X\circ \mathrm{C}\left(I, \mathrm{S}^*_{\langle t\rangle} \circ \mathrm{C}\left(\alpha_Y\circ A, \alpha_Y\circ \mathrm{S}\left(B,C\right)\right)\right)$
}\label{fig:label1}
\end{figure}

To illustrate, consider
for example the application in \cref{fig:label1}.
 Consider then a run-time observation 
of an activity related to the inner $B$ network instantiated from tag $\langle t \rangle= 123$ in the replication selection.
When looked up from label $X$, the functional network index is $[1;123;1;0]$; when looked up from
label $Y$, the network index is simply $[0]$. 
\item
\emph{arity inspection}, which given a network index approximates the activity/liveness/agent
arity of the replica identified by $n$.

To illustrate, consider the example from \cref{fig:label1} after
$\langle t \rangle$ has run through the network with values 123 and
124. In a single instance of $X$, the liveness arity for both 
sub-networks identified by $Y$ is 2.
$A$ has two live replicas $[1;123;0]$ and $[1;124;0]$; whereas
$\mathrm{S}(B,C)$ has replicas $[1;123;1], [1;124;1]$.

\item
\emph{state inspection}, which given a network index enumerates all
the inner constructs that currently have state and the type of state
they hold. Active transducers are listed with the name of the state
they are currently in and the content of their currently full hold
variables; active reorderings ($\mathrm{R}$) with the number of
records currently held; replications with the number of replicas
currently maintained, etc.
\end{itemize}

The lookup service is akin to the reverse lookup service commonly found
in debuggers. The arity and state inspection services are intended to provide
a handle on the current management state maintained by the coordination layer.

\section{Relationship to \snet{} and design rationales}\label{sec:relsnet}
\label{sec:relsnetfun}
\label{sec:relsnetefun}
\label{sec:efunmotiv}

\subsection{Overview of functional changes to \snet{}}

\spnet{} uses the same streaming foundations as \snet{} and reuses its
primitive networks for boxes and filters. There are however three
major functional updates in \spnet{}, motivated by practical
experiences using \snet{}.  As we discuss in \cref{sec:stmotiv},
\spnet{} introduces exact type match and dedicated channels via type
tags where \snet{} uses mostly structural record subtyping.  As we
motivate in \cref{sec:syncmotiv}, \spnet{} merges \snet{}'s filters
and synchrocells into the single finite state machine construct called
transducer.  \spnet{} furthermore introduces unordered replication
composition, as motivated in \cref{sec:aumotiv}.

Next to these major updates, a few aspects of \snet{} have been
clarified and simplified in \spnet{}.  \spnet{} makes reordering
orthogonal to other functional composites via its reordering
combinator, while \snet{} provides separate variants of combinators
that preserve input order.  While \snet{} solves the routing ambiguity
in its selection combinator ``non-deterministically'' without stating
what strategy is actually used, \spnet{} specifies it; any
non-determinism, if so desired, can be expressed \emph{and controlled}
using extra-functional selection (\cf \cref{sec:ersel}).

\subsection{Stream connections and structural typing}
\label{sec:stmotiv}

Both \snet{} and \spnet{} are based on the transformation of a general
acyclic application graph into a serial-parallel structure with
bypasses for non-matching types. Support for cyclicity was added in
\snet{} via replication composition. The overall design seems to work
very well with most of our industrial use cases. It is not free,
however, from certain design drawbacks.

First of all, the serial-parallel representation, \ie a pipeline of
groups of boxes connected in parallel, generally requires a bypassing
mechanism as the original transformation suggested. To understand this
requirement, one can think of a node of the graph sending directly to
another node, whose maximum distance from the sources of the graph is
greater by more than one than that of the sending node. After the
transformation those nodes will be separated by one or more pipeline
stages that would have to be bypassed. The original \snet{} design did
not appreciate the ubiquity of bypasses in any real program, even
though special compact notation (\texttt{->}) was included to avoid
long-handed specifications.

From the bulk of experience that we have accumulated over the recent
years it does appear that the remedy should come from the type
system. Indeed, already (and according to some, rather inelegantly)
tags in \snet{} are classified into ordinary and binding, the latter category
intended for pruning the flow inheritance tree that results from
combining input types when the most general acceptable set of records
is being determined. Binding tags are indeed not only necessary but
they should be seen as ``normal'' as they are the only ones that can
separate variants. For example a network accepting $\{a\}$,
$\{b\}$ and $\{c\}$ as three variants and one which
has a response to each of those would indeed have to have a response
to all seven distinct nonempty unions of these sets as various
subtypes of the originals. Worse still, all but the singleton sets of
these unions would lead to a nondetrministic choice in identifying
which original variant they should be assigned to, with the extra
members of the set being flow-inherited. Unfortunately, explicitly
separating the variants with tags in the spirit of algebraic data
types as well as standard practice of imperative programming languages
does not solve the problem: the variants $\{\langle i\rangle, a\}$,
$\{\langle ii\rangle, b\}$ and $\{\langle iii\rangle, c\}$ are as prone to the
undesired inheritance as the original ones because the tags, too, can
be inherited unless they are binding. Consequently in this example
only $\{\langle \#i\rangle, a\}$, $\{\langle \#ii\rangle, b\}$ and
$\{\langle \#iii\rangle, c\}$ represent a solution that is free of spurious
subtyping, as it only accepts records that contain either a or b or c
as befits a proper variant record type.

In the light of this, we found desirable to simplify the typing
rules of \snet{} and promote a single form of tags that is always binding, and
explicitly defines the end-points of streams. Such tags also
inhibit inheritance but in a different way: while \snet{}'s binding tags
demand that the receiving end of the match offers the same tag or else
this would be a type error, when \spnet{}'s new tags fail to match they cause
the record they belong to to be bypassed over the unmatched
entity. \spnet{}'s tags delay a type error, rather than merely
causing it to be ignored. If a record carrying a tag reaches the exit
of a net environment, or fails to match an explicit network type
signature, a type error does result.

\subsection{Usability of synchrocells}\label{sec:syncmotiv}

Another problem was that \snet{} does not have a satisfactory solution for
is the variety of synchronisers that are required in the real
world. Its simple $n$-ary synchrocells have a finite and rather trivial
state machine behaviour. The idea behind them was originally that any
behaviour more complex than that would be realised via unfolding
replicative structures that contain synchrocells and boxes that compute
transitions between various states of the required compound
synchroniser. For instance if the application needs a synchrostructure
that combines records of type 1 with some persistent record value of
type 2, which is also periodically updated by records of type 2, such
a structure could be realised as a network where records of type 2
cause synchronisation in a synchro-queue while at the same time
duplicating themselves to be send to the next stage of the queue thus
mimicking real persistency. Unfortunately, this method is not just
awkward, it also requires a really inelegant action when the
persistent values are to be modified. At such moments, a spurious
record of type 1 would need to be produced only to be discarded at the
exit of the network (for which a tag should be flow-inherited to
indicate that the record is spurious). These regrettable manoeuvres
seem unavoidable in the original vision while at the same time they
are also illogical as far as the basic principle is concerned. The
principle was to divide the boxes into ``can compute, have no state''
and ``have a state, cannot compute'', the latter in the sense of
modifying box values. This principle does not exclude synchronisers
with complex behaviour as long as all they do is select what to store
and what to pass on. The original \snet{} synchrocells seem as
undeservedly specialised for one kind of synchronisation task as would
boxes with only one output variant be for one task of computing. To
make matters worse, the behaviour described above is very plausible:
any application that wishes to tune the parameters of its algorithm
from time to time would need the above kind of synchroniser as a
building block.

The solution lies in introducing a synchro-construction language
similar to the filter language from \snet{}. Such a language
can be simple and elegant, allowing the coordination programmer to
specify state transitions between the synchroniser states as well as
output and stored values in terms of set operations on input and
stored records. The state transitions should logically be allowed to
depend on values found in input records. It is easy to satisfy oneself that
the corresponding transducer language in \spnet{} is not much
larger than the filter and synchrocell languages from \snet{}. Besides,
as soon as this language exists, it becomes possible to express
stateless synchronizers that perform simple tasks on input records,
and the filter construct from \snet{} can be dropped in favor
of the more general construct.

\subsection{Aggregate updates}\label{sec:aumotiv}

The data flow model on which \snet{} is built requires complete
encapsulation: all that a box can read is its input record and all
that it is supposed to write into is output records. This has worked
surprisingly well in most use cases. However, when the application
requires parallel processing of a large shared datastructure, such as
a large graph or a distributed database, the basic assumption of
dataflow could easily become burdensome. What is required is a
slightly different discipline: sharing should be allowed but the box
must yield to the coordinator's control when it attempts
sharing. Specifically, we anticipate the need to access a large shared
data structure from a box and only to read and modify a small part of
it; we expect that such modifications done by different boxes collide
infrequently and that when they do a transaction discipline must be
enforced to preserve serialisability as its essential semantics. In
other words a box should be able to access the shared object, read a
small part of it and modify a small part of it and then terminate (as
boxes usually do), without any other box modifying the data being read
by the first box before it terminates. The ordering of such
modifications would be linear but nondeterministic, which in practice
means optimistic parallelism with failures and retractions.

In terms of \snet{} language constructs, we are lacking a combinator
which would connect replicas of its operand serially in
non-deterministic order and stream the data structure through them,
while at the same time delivering parameter records to the replicas in
a parallel fashion: a non-deterministic parallel-serial
combinator. This combinator is introduced as unordered replication
composition in \spnet{}.

\subsection{Identification of run-time activities}

Perhaps the most painful shortcoming of \snet{} while using it in practice
is the lack of insight during application execution to relate
a run-time behavior back to its origin in the application specification.

For example, a common situation is to detect that a particular
procedure in the \snet{} application's binary code concentrates most
of the execution time, but only for some of its activations. This
situation can appear when a component box has different behavior
classes depending on the data provided as input. The question then
appears: in which path through the application network is this
behavior class triggered?

A textual locus in the static application specification is usually not
sufficient to isolate the origin of a behavior. With replication, a
single component may be instantiated multiple times during execution,
with all replicas executed concurrently. In general, the run-time
identifier of a network replica is necessary to identify the cause of
an observation.
Another realization is that not only component code is worthy of the
optimization engineer's attention. In~\cite{mckenzie.13.fdcoma}, the
authors demonstrate that internal \snet{} management tasks were the
cause of a load imbalance in a parallel computation. Unfortunately, in
contrast to component boxes all the internal tasks of \snet{} are
anonymous and cannot be easily related to the input application
specification.

What these scenarios have revealed is that \spnet{}, like any
programming environment, needs a ``reverse lookup'' mechanism from
space/time loci during execution back to a specification's source
code. However, a simple mechanism based on a static mapping of memory
addresses to source locations, like those used by traditional
debuggers, would be insufficient for a highly concurrent environment
where multiple processes running at the same address may emerge from
multiple source locations or paths through the
specification. Consequently, we introduce a naming scheme with
\spnet{} (\cref{sec:rtident}), a new labeling combinator $\alpha$ and
selection scheme (\cref{sec:netlabel}), and run-time services to
  inspect a running application (\cref{sec:implsvc}).

\subsection{Entity-centric \vs record-centric projections}
\label{sec:projmotiv}

When \snet{} was originally designed, its language operators
were constructed as abstractions of common patterns
when engineering applications made of concurrent processes.
The unphrased assumption behind the design of \snet{} was
that entities in a specification should abstract run-time
processes, and conversely that each entity would be reifed during execution using
a process; and thys that inter-component streams would be naturally
implemented as buffered channels connecting the processes together. In
this vision, a run-time execution of an \snet{} application defines
processes for each box, plus additional control processes
for the composition operators: ``splitter'' processes before
parallel composition to redirect records to the sub-processes
depending on their type, ``merger'' processes after composition to
restore order, ``synchronizer'' processes for synchrocells, etc.

Meanwhile, the functional definition of \snet{} did not altogether
mandate that an implementation must use this \emph{projection} onto
process networks to actually carry out the application's execution.
This under-specification led multiple researchers to realize later on,
separately, that there exist other possible run-time projections of an
\snet{} specification.


\begin{figure}
\centering
\subfloat[Entity-centric execution.]{\includegraphics[scale=.4]{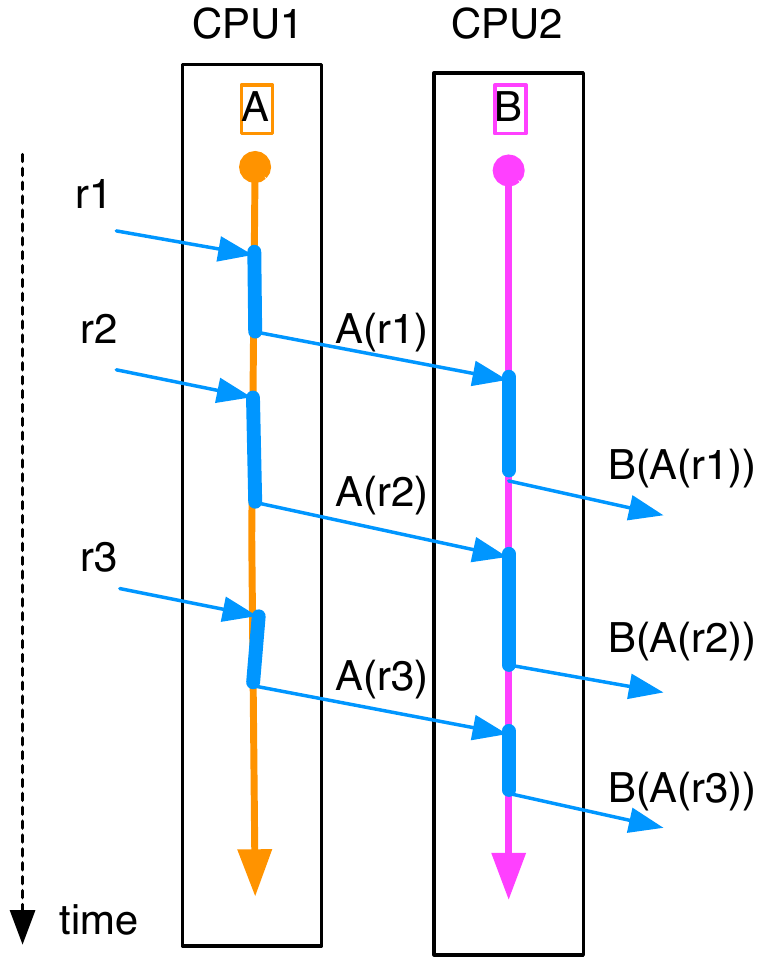}\label{fig:abent}}
\quad
\subfloat[Record-centric execution.]{\includegraphics[scale=.4]{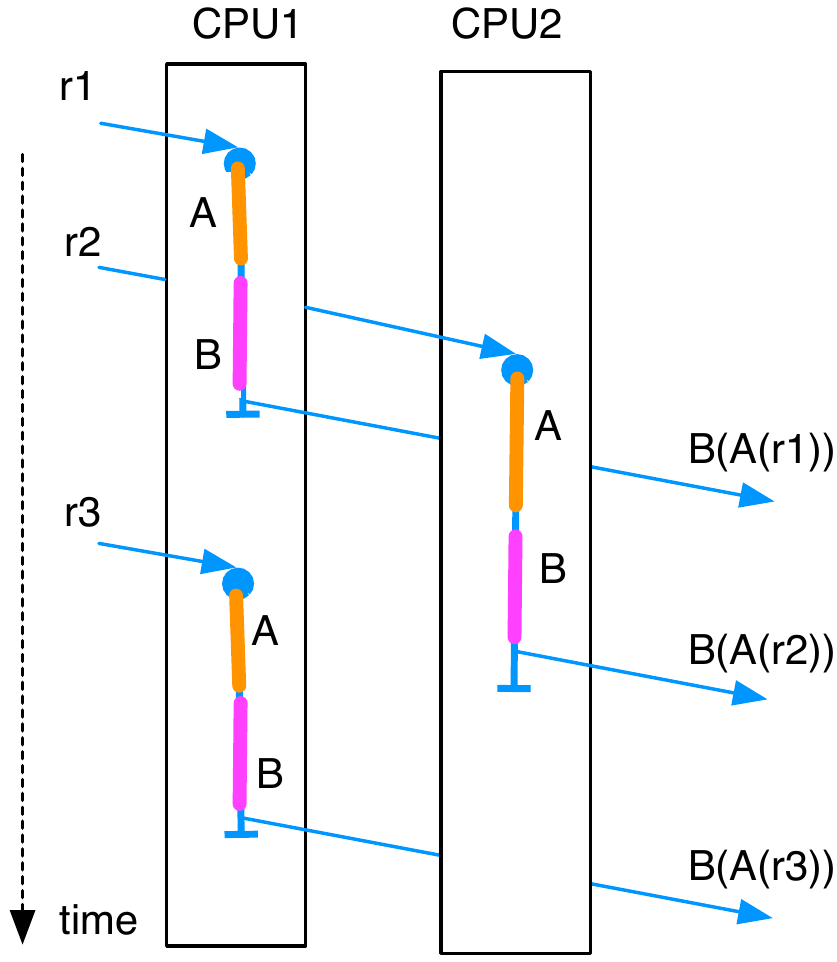}\label{fig:abrec}}
\caption{Two valid run-time projections of \texttt{A..B} onto processes.}\label{fig:abproj}
\end{figure}

One such approach can be found in~\cite{holzenspies.10}, where the author
suggests to create processes for each individual \emph{record},
instead of box or control entity. Each record-process then computes
sequentially and recursively the transformation operated for that
input record over the length of the \snet{} application
specification. To illustrate this approach, we can consider for
example the pipeline of A and B in sequence, \ie \texttt{A..B}. The ``natural''
projection of this network onto processes is given in
\cref{fig:abent}: one process is created to compute transformations of
inputs by function A; another process is created to compute
transformations by B. The ``transposed,'' record-centric projection is
given in \cref{fig:abrec}: one process is created for each input; each
process sequentially computes A then B for this input, then
terminates. It is easy to see that both projections are
\emph{functionally equivalent}, in that they compute the same
relationship between input and output according to the \snet{}
semantics.

In a record-centric vision, stream synchronization can be implemented by
joining concurrent threads; boxes with multiplicity by creating new streams
for any new additional record injected in the network. The reader
is referred to~\cite[Chap.~7]{holzenspies.10} for a detailed description
of the mechanisms involved.
The existence of multiple valid projections of \snet{} was quickly
recognized as more than a mere intellectual curiosity. Indeed, the
projection that maximizes throughput or reduces latency for a given
execution platform \emph{depends on the platform's parameters}
and the algorithmic complexity of the boxes themselves.

An entity-centric projection is desirable for compute-bound
applications, when the objective is to specialize processors to the
component's function in order to accelerate them. However, an
entity-centric projection pays a price in locality: the intermediate
results between components travel spatially and internal communication
costs are increased. If the platform does not provide sufficient
bandwidth between processors, this projection quickly becomes
communication-bound. In practice, we have seen this happen
when the computational complexity of components was not enough
to mask the limited memory bandwidth available to multiple cores in
a shared-memory system, or when attempting to map \snet{} applications
to platforms with GPU accelerators.

Conversely, a record-centric projection is desirable for
applications with large communication requirements on their internal
streams compared to the requirement on their input and output streams,
\ie when the objective is to maximize locality. A record-centric
projection pays a price in efficiency. Each processor must be
sufficiently general to perform all the transformations in the
application, and code locality can become an issue. Moreover,
synchronization state is likely to be shared across arbitrary
processors in the system, and synchronization operations may thus have
higher latencies.

The existence of this duality and its consequences w.r.t optimization
opportunities suggests to equip \spnet{} with mechanisms to let an
optimization engineer select which type of execution projection to use,
separately from the functional specification of the
application. This is introduced via the $\gamma$ combinator (\cf \cref{sec:projs}).

\subsection{Lifetime of activities}\label{sec:taumotiv}

The most recent entity-centric implementation of \snet{}, currently
used as research vehicle, uses cooperative scheduling of lightweight
tasks over worker threads pinned to hardware
processors~\cite{prokesch.10}. Cooperative scheduling lowers the cost
of concurrency management overall by avoiding context switches to the
operating system and maintaining separate task state and scheduling
queues per worker. However, cooperative scheduling, by construction,
places the scheduling responsibility in the hands of the tasks
themselves: once started by a worker scheduler, it stays assigned to
that worker until it terminates. Extending the implementation to allow
dynamic task migration while a box-task is transforming records is
undesirable for two reasons: it would require to re-introduce the
overhead of preemption, as well as mutual exclusion for task
management state between workers, which the lightweight task/worker
infrastructure was intended to avoid in the first place.
In contrast, a mapping decision prior to a task's creation is
cheap to implement since there is no state yet to synchronize nor
activity to preempt. Intuitively, an application's execution where
tasks are short-lived thus offers more opportunities for cheap dynamic
load balancing than a system where tasks persist after their creation.

\begin{figure}
\centering
\includegraphics[scale=.4]{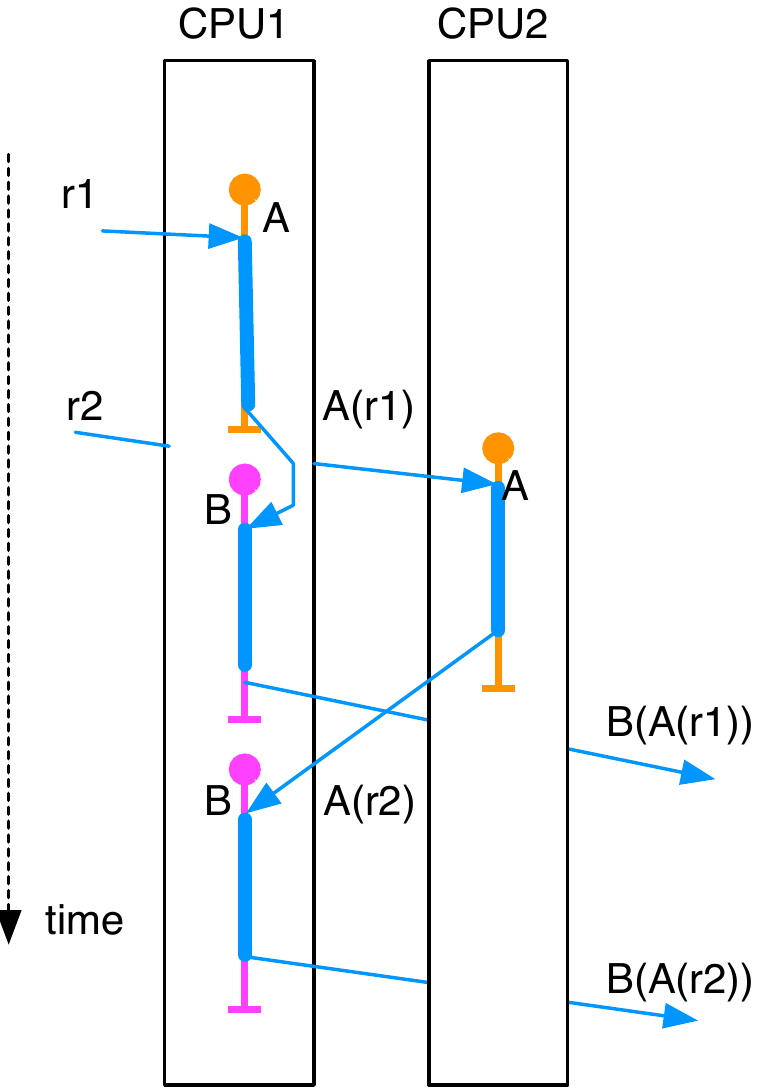}
\caption{Short-lived entity-centric tasks for A..B.}\label{fig:abrepl}
\end{figure}

This is the starting point of a modified implementation of \snet{}
that was finalized in the second half of 2012: instead of projecting
\snet{} entities into long-lived tasks that repeatedly read from their input
stream and process one input record, in a loop (\cref{fig:abent}), this
implementation causes each entity-task to terminate as soon as one
record is processed, so as to allow a new mapping decision to be made
for the next input record (\cref{fig:abrepl}).

In other words, by shortening the \emph{lifetime} to a task to the
minimal amount of computation, that is the handling of exactly one
input record by a box, we obtain the maximum flexibility for
scheduling short of re-introducing fully-fledged preemption. However,
this comes at a cost: the number of tasks created and cleaned up by
unit of time increases, which amounts to re-introducing some
management overhead. For homogeneous workloads this overhead is as
undesirable as it is unnecessary, since homogeneity implies that
fine-grained scheduling decisions have a low impact on overall
execution time.

While we have understood these trade-offs in the specific context of
cooperative scheduling of tasks over worker threads, this context is
merely an instance of a more general, fundamental aspect of
concurrency management. Regardless of the mechanism used to implement
concurrency over hardware processors, the chosen granularity for
non-preempted, non-migrated activities determines a trade-off between
throughput and jitter. Furthermore, there does not seem yet to exist
any consensus about the best way to detect the optimal granularity for
a given workload/platform combination. In particular, any approach
which sets global values (\eg the \texttt{HZ} property that sets the
time slice duration on Unix systems, commonly set to 1 or 20ms) is
typically inappropriate in the light of heterogeneous applications or
resources where the granularity should change depending on the
application part, possibly dynamically.

With \spnet{}, we thus propose to explore this trade-off using a novel
approach based on two principles. First, we state that a computing
agent (either a task, thread, or any other mechanism used to perform
an elementary computation on the platform at hand) is assigned to
hardware resources upon its creations and is not migrated throughout
its lifetime.  This assignment can be guided with the $\phi$
combinator (\cf \cref{sec:phi}).  Then, we establish that each
transformation specification in the \spnet{} abstraction can be
projected not onto only one, but multiple successive agents during
execution; this granularity can be in turn controlled via the $\tau$
combinator (\cf \cref{sec:projs}). This way, the throughput/jitter
trade-off can be controlled flexibly for each sub-network in an
application.

\subsection{Environmental awareness}

The practical application of \snet{} has revealed consistently that
some tuning parameters in an application specification are highly
dependent on the execution resources actually available at run-time.

For example, a common pattern found in compute-bound applications is
the divide-and-conquer network: a ``splitter'' box splits large input
records into sub-streams of smaller records to be processed
concurrently by parallel ``compute'' boxes. Users of this pattern
typically face a challenge: how to decide the size of the
smaller workloads?

In this example, there are two main scenarios.
When the workload is heterogeneous, an incentive exists to split the
workload in smaller sizes to give the coordination layer more
opportunities to balance load dynamically across hardware
resources. However, as the granularity becomes finer, the overheads of
coordination increase relative to computation time; after a threshold
the coordination overheads dominate and the throughput is actually
reduced. Of course, this threshold depends both on the application
\emph{and the execution platform}: any specification-time granularity
choice must be revised after profiling.

When the workload is homogeneous, an incentive exists to spread the
workload evenly across the available hardware resources. The
granularity should be as coarse as possible to minimize coordination
overheads while obtaining the maximum parallelization possible on the
platform.  However, two obstacles prevent an optimal
specification-time choice. For one, the size of memory caches in the
execution platform determines a threshold past which cache thrashing
will actually reduce throughput. Second, if more concurrent
transformations are defined than there are processors
available at run-time, context switch overheads also reduce throughput
scalability.

This example is only one instance of a need for an \emph{environmental
  feed-back loop}, whereby the coordination engineer can specify
tuning parameters, such as sub-workload granularity selection in the
example, as a function of the resources actually available on the
execution platform. In \spnet{}, we propose to introduce such
feed-back loops via the primitive network $\delta$ (\cf \cref{sec:envaware}) which observes
the execution environment and injects observations as scalar values
that can be used in the application's coordination logic. These
can be then combined with $\rho$, $\theta$ and $\phi$ to tune
the application's behavior according to extra-functional requirements and budgets.

\section{Summary and conclusions}\label{sec:conc}

This report has presented the design of \spnet{}, a declarative
language for describing networks of asynchronous components and their
execution semantics in resource-contrained execution
environments. While it reuses concepts from its predecessor \snet{},
\spnet{} simplifies \snet{}'s set of functional combinators and adds a
more comprehensive synchronization facility, the transducer. In the
extra-functional domain, \spnet{} provides facilities for:
\begin{itemize}
\item naming of networks and non-local name references to coordinate
  behavior;
\item environment awareness, to react to resource
  availability;
\item fault tolerance by enabling explicit exception handling;
\item constraining the operational semantics of arbitrary sub-networks towards either a process-centric
  projection or a record-centric projection;
\item contraining the mapping of networks to processing resources;
\item constraining the execution to satisfy resource budgets.
\end{itemize}

The definition of \spnet{} is an effort concurrent to the definition
of AstraKahn, another coordination language by the same research
groups.
The main difference between \spnet{} and AstraKahn is that
all valid \spnet{} programs are serializable, whereas valid AstraKahn
programs may contain non-serializable network cycles.

\section*{Acknowledgements}
\addcontentsline{toc}{section}{Acknowledgements}

This research was partly supported by the European Union under grant number
FP7-248828 ADVANCE\footnote{\url{http://www.project-advance.eu/}}. The
authors would like to thank the ADVANCE partners for the constructive
criticism of \snet{} that has resulted in \spnet{}.

\newcommand{\etalchar}[1]{#1} 
\addcontentsline{toc}{section}{References}
\bibliographystyle{is-plainurl}
\bibliography{doc}

\end{document}